\documentclass[aps,10pt]{revtex4}
\usepackage{epsfig,graphics,amssymb,amsmath,subeqnarray}
\def\r{{\bf r}}\def\rp{{\bf r}_\perp}\def\di{\displaystyle}
\def\d{{\rm d}}\def\p{\partial}\def\u{{\bf u}}\def\I{{\cal I}}
\def\L{L}\def\cpe{\xi_{\perp}}\def\g{\gamma_\perp}
\def\cpa{\xi_{\parallel}}\def\t{{\bf t}}\def\n{{\bf n}}
\def\RFUpa{{\cal R}_{\parallel}^{FU}}\def\RFUpe{{\cal R}_{\perp}^{FU}}
\def\RLOpe{{\cal R}_{\perp}^{L\Omega}}
\def\RFUxx{{\cal R}_{\parallel}^{FU}}\def\RFUyy{{\cal R}_{\perp}^{FU}}
\def\RFUzz{{\cal R}_{\perp}^{FU}}\def\RLOxx{{\cal R}_{\parallel}^{L\Omega}}
\def\RLOyy{{\cal R}_{\perp}^{L\Omega}}\def\RLOzz{{\cal R}_{\perp}^{L\Omega}}
\def\dyx{\frac{\p y}{\p x}}\def\dzx{\frac{\p z}{\p x}}
\def\ddyx{\frac{\p^2 y}{\p x^2}}\def\dddyx{\frac{\p^3 y}{\p x^3}}
\def\ddzx{\frac{\p^2 z}{\p x^2}}\def\dddzx{\frac{\p^3 z}{\p x^3}}
\def\dyt{\frac{\p y}{\p t}}\def\dzt{\frac{\p z}{\p t}}
\def\Ox{\Omega_x}\def\Oy{\Omega_y}\def\Oz{\Omega_z}

\begin{document}

\title{Floppy swimming: Viscous locomotion of  actuated elastica}
\author{Eric Lauga\footnote{Email: lauga@mit.edu}}
\affiliation{Department of Mathematics,
Massachusetts Institute of Technology,
77 Massachusetts Avenue,
Cambridge, MA 02139.}
\date{\today}
\begin{abstract}
Actuating periodically an elastic filament in a viscous liquid generally breaks the constraints of Purcell's scallop theorem,  resulting in the generation of a net propulsive force. This observation suggests a method to design simple swimming devices - which we call ``elastic swimmers"  - where the actuation mechanism is embedded in a solid body and the resulting swimmer is free to move. In this paper, we study theoretically the kinematics of elastic swimming. After discussing the basic physical picture of the phenomenon and the expected scaling relationships, we derive analytically the elastic swimming velocities in the limit of small actuation amplitude. The emphasis is on the coupling between the two unknowns of the problems - namely the shape of the elastic filament and the swimming kinematics -  which have to be solved simultaneously. We then compute the performance of the resulting swimming device, and its dependance on geometry. The optimal actuation frequency and  body shapes  are derived and a discussion of filament shapes and internal torques is presented. Swimming using multiple elastic filaments is discussed, and simple strategies are presented which result in straight swimming trajectories. Finally, we compare the performance of elastic swimming with that of swimming microorganisms.

\end{abstract}
\maketitle

\section{Introduction}
The fluid mechanics of microorganism locomotion, pioneered more than fifty years ago by G.I. Taylor, has become one of the most successful branches of biomechanics, with success in both the basic  physical understanding of  flow behavior  and the quantitative prediction of kinematics and energetics of locomotion  \cite{taylor51,taylor52,gray55,lighthill76,brennen77,purcell77,bergbook,braybook}. 

Recent technical advances have led to ever more precise fabrication at small scales (microns or less), prompting both theorists \cite{becker03,avron04:opt,najafi04,najafi05,avron05:pushme} and experimentalists \cite{dreyfus05} to design and analyze a series of simple low-Reynolds number swimmers. The experiment of Dreyfus {\it et al.} \cite{dreyfus05}, in particular, reported locomotion in  a sperm-like micro-swimmer, composed of a cargo (red blood cell) and a slender flexible filament made of a series of paramagnetic beads. In that case, actuation by oscillating transverse magnetic fields led to the generation of  bending waves propagating along the filament and resulted in the motion of the micro-swimmer.  In this system, the right-left symmetry was broken by the presence of a cargo and led to a preferential tip-to-base propagation of the bending waves, resulting in locomotion in the direction base-to-tip.

An alternative way to break the symmetry in a similar system would be to build-in the asymmetry in the actuation. In particular, if an elastic filament is periodically actuated at one of its extremities in a viscous liquid, the resulting motion will lead to the propagation of bending waves and, in general, propulsive forces. This idea was originally proposed by Edward Purcell \cite{purcell77}. Physically, actuating an elastic filament allows one to break the constraints of the ``scallop theorem'' - which states that a body performing a reciprocal motion at low Reynolds number cannot propel itself - by allowing the boundary conditions on the fluid problem - that is, the shape of the filament - to be itself a function of  the fluid flow. The original theoretical study on this problem was proposed by Wiggins and Goldstein  \cite{WigginsGoldstein}, who showed that the amplitude of the actuated elastic filament satisfies a  hyperdiffusion equation. This equation was also derived in earlier work by Machin in the context of wave propagation in the flagella of swimming microorganisms \cite{machin58,machin63}. A similar theoretical treatment was proposed as a simple model of  the sliding filament model of eukaryotic axonemal beating by coupling the elasto-hydrodynamics problem  with models for the  behavior of active molecular motors \cite{camalet99:prl,camalet00:njp}. 

The main features of this problem have been successfully exploited experimentally to measure the bending modulus of biopolymers (actin filaments  and microtubules), either using thermal fluctuations \cite{gittes93} or using an active actuation \cite{riveline97,Wiggins:Biophys}. Related studies include the dynamics of magnetic filaments \cite{cebers05,roper06,gauger06},  the three-dimensional actuation and instabilities of flexible filaments \cite{wolgemuth00,powers02,manghi06,wada06} and the exploitation of symmetry-breaking to pump fluid in a channel \cite{kim06:pumping}.

In this paper, we consider the case where the actuated flexible filaments are exploited for locomotion purposes. We consider a prototypical micro-swimmer composed of a solid body and an elastic slender filament (see Fig.~\ref{mainfig}). The filament is fixed to the body and its base-angle is varied sinusoidally at frequency $\omega$, in two or three dimensions,  by a mechanism embedded in the swimmer body. This generates the propagation of bending waves down the elastic filament and propels the swimmer forward.  This design for an ``elastic swimmer" is the simplest locally-forced low-Reynolds swimmer exploiting the interplay of fluid drag and bending rigidity for propulsive purposes, and as a result, its swimming kinematics and energetics are of fundamental interest. 

As we will see below, the locomotion of the elastic swimmer  also turns out to be an interesting  mathematical problem. Indeed, to characterize the swimmer completely, two main problems need to be solved for, namely (1)  the periodic shape of the elastic filament  and (2)  the kinematics of swimming. However, these two problems cannot be solved independently. Along the filament, drag forces and bending forces balance. On one hand, the swimming kinematics affects the drag and therefore the filament shape. On the other hand,  the shape influences the viscous propulsive force and therefore the overall swimming velocity. As a result, these two problems have to be solved simultaneously, a fact which results - as will be seen below -  in the appearance of integro-differential equations.  This feature might have been overlooked by previous analytical studies.

Numerical simulations of the elastic swimming problem were presented by  Lagomarsino {\it et al.} \cite{lagomarsino03}  using particle-based  methods  (see also Ref. \cite{lowe03}). However, in this study, the filament was actuated by  external forces and torques, and as a result is fundamentally different  from the self-contained force-free and torque-free swimmer which we consider in this paper. Moreover, in the case of small amplitude oscillations of the actuation point,  the simulations by  Lagomarsino {\it et al.}  obtain a constant swimming velocity for long filaments (long in the sense  $L\gg \ell_\omega $, see below), whereas in fact the velocity should decrease to zero because of excessive drag on the filament. This discrepancy is resolved in our paper.

Numerical simulations of the three-dimensional actuation (rotation) of the filament were presented by Manghi {\it et al.} \cite{manghi06} using particle-based methods which include hydrodynamic interactions (similar to those used to study polymer dynamics). However, the simulations by Manghi {\it et al.} obtain swimming even in the case where the body sizes shrink to zero, a fact which also violates torque balance for a torque-free swimmer at zero Reynolds number \cite{chwang71,keller76}.

Recently, Yu {\it et al.} has performed a macro-scale experiment aimed at measuring the propulsive force generated by   actuated elastic filaments in Stokes flows and compared it with existing theories  \cite{yu06}. The filaments were fixed in space with base-angles which were actuated sinusoidally and  very good quantitative agreement was found between the measured propulsive force and that predicted by the small-amplitude theoretical study of Wiggins and Goldstein \cite{WigginsGoldstein}.

This paper is organized as follows. In \S\ref{pp} we present the basic physical picture for the coupling of hydrodynamics and bending forces in actuated filaments. We estimate the optimal actuation conditions of the filament and derive the expected scalings for the swimming speed of the swimmers. In \S\ref{main} we derive the swimming kinematics of the elastic swimmers analytically in the limit of small actuation amplitude. The assumptions necessary to perform the calculation are clearly stated, and the final results are six analytical formulae for the three-dimensional trajectory of the swimmer (Eq.~\ref{final_formulae}). The performance of the elastic swimmer is discussed in \S\ref{S:optimal}. In particular, we characterize optimal swimmers as well as the difference between filament shapes for free-swimming versus  fixed  actuated filaments. Elastic swimming with more than one flexible filament is discussed in 
\S\ref{S:many} and we show that steady swimming on a straight line can be obtained with six filaments.
Finally, a discussion of the results and a comparison of the swimmer performance with swimming microorganisms are presented in \S\ref{discussion}.

\section{Physical picture}
\label{pp}
\subsection{Elasto-hydrodynamics}

\begin{figure}[t]
\begin{center}
\includegraphics[width=.95\textwidth]{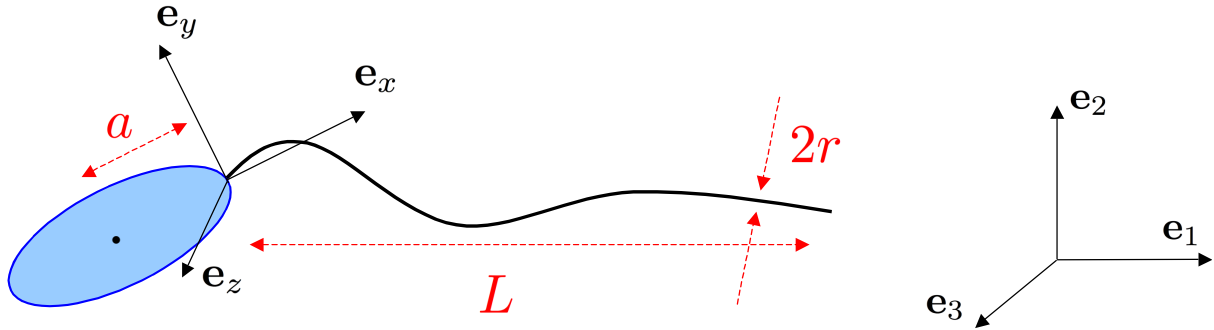}
\caption{Notations for the elastic swimmer. The filament has length $L$ and radius $r$. The distance between the center of mass of the body and the actuation point is denoted $a$. The frame $\{{\bf e}_x,{\bf e}_y,{\bf e}_z\}$ moves with the swimmer body whereas the frame $\{{\bf e}_1,{\bf e}_2,{\bf e}_3\}$ is fixed. }
\label{mainfig}
\end{center}
\end{figure}

As explained above, we consider in this paper the prototypical elastic swimmer displayed in Fig.~\ref{mainfig}.
We denote by $L$ the length of the filament, $r$ its radius, $A$ its bending stiffness, and $\xi_\perp$ its normal  drag coefficient {\it i.e.} the viscous force exerted by the fluid per unit length of the filament for motion perpendicular to its length \cite{gray55,lighthill76,brennen77}. Let us first consider the case where the filament is actuated but not free to move \cite{WigginsGoldstein,yu06} . If $y$ denotes the typical amplitude of  a material point at a distance $x$ along the filament, the balance between local viscous drag and bending forces on the filament results in a hyperdiffusion equation for small-amplitude motion,
\begin{equation}
\label{hyperdiff}
\xi_\perp \frac{\p y}{\p t} = - A\frac{ \p^4 y}{\p x^4}\cdot
\end{equation}
For a given actuation frequency $\omega$, inspection of Eq.~\eqref{hyperdiff} shows the appearance of an intrinsic bending-hydrodynamics length scale, $\ell_\omega = (A/\omega \xi_\perp )^{1/4}$ \cite{WigginsGoldstein}. If $L\ll \ell_\omega$, bending forces win and the filament is straight. On the contrary, if $L\gg \ell_\omega$ drag forces win and the portions of the filament located at a distance larger than $\ell_\omega$ from the actuation point are essentially straight, a feature which is expressed mathematically by an exponentially decay of the amplitude of the solution to the elasto-hydrodynamics problem over the length scale $\ell_\omega$ \cite{WigginsGoldstein}. 

In the case where the actuated filament is embedded in a swimming device, the limit  $L\ll \ell_\omega$ results in a reciprocal actuation of the filament, and therefore - by use of the scallop theorem - the swimming velocity is zero  \cite{purcell77}. The other limit $L\gg \ell_\omega$ leads to a constant value of the propulsive force, as there are no contribution from any portion of the filament beyond  $L\sim \ell_\omega$, and since in this case the  viscous drag is large, and this limit   also results in swimming velocity decreasing to zero. As a consequence, we expect that the optimal swimming  will be obtained for a filament length  $L\sim\ell_\omega$, taking full advantage of the drag-induced bending of the filament while keeping the overall drag on the swimmer  low \cite{lowe03,lagomarsino03}. This result is confirmed in \S\ref{S:optimal} where we compute the optimal elastic swimmers.

\subsection{Optimal forcing}

Before considering the swimming kinematics, we show here using scaling arguments that  it is energetically favorable, when actuating the elastic filament periodically, to only use the fundamental frequency. Let us call $T$ the  actuation  period and  $\epsilon$ its amplitude. Such periodic actuation is necessary in order to constantly generate bending deformation and therefore propulsion. The mechanical forcing at the base of the filament can, in principle, include the fundamental frequency $2\pi/T$, and all other harmonics. For small-amplitude motion, the  propulsive force is quadratic in the filament dynamics, so by orthogonality we can study each frequency independently.
Let us consider forcing at a given frequency $\omega$. The elastic propulsive force scales as \cite{WigginsGoldstein,yu06}
\begin{equation}\label{prop}
F\sim A \int_0^L \frac{\p y}{\p x} \frac{\p^4 y }{\p x^4}\,\d x .
\end{equation}
For an optimal propulsion, the filament length $L$ is on the order of the intrinsic length scale $\ell_\omega$. Since we have $y\sim \epsilon \ell_\omega$, we expect   Eq.~\eqref{prop} to scale  as $F\sim A \epsilon ^2 / \ell_\omega^2\sim \epsilon^2 (A  \xi \omega)^{1/2}$, where $\xi$ is the typical value of the drag coefficient. 
As a difference, the total work done by the actuator against the viscous fluid in the period $T$ is given by
\begin{equation}
W\sim T  \int_0^L \xi \left(\frac{\p y}{\p t}\right)^2 \,\d x,
\end{equation}
which scales  as $W\sim \epsilon ^2 T  \xi   \ell_\omega^3  \omega^2\sim \epsilon ^2 T (A^3\xi\omega^5)^{1/4}$. To within logarithmic terms (arising in $\xi$, see below), the propulsive force scales therefore with frequency as $F \propto \omega^{1/2}$ whereas the total work done scales with a higher power as $W\propto \omega^{5/4}$. For a given period and a given amount of energy available, the maximum propulsive force will therefore be obtained by actuating the filament only at the minimum frequency possible, $\omega = 2 \pi / T$, with no harmonics.

\subsection{Swimming}
We now consider the swimming kinematics and derive here the expected scaling for the mean swimming velocity. The detailed calculations will be presented in the main section of the paper. We assume that $L\sim \ell_\omega$ as this is the case where the optimal propulsion is expected to be generated, and we fix the amplitude, $\epsilon$, and frequency, $\omega$, of the actuation. 

First, we consider the case where the body, of typical size $a$, is much larger than the filament length, {\it i.e.} $a\gg \ell_\omega$. In that case, the large viscous resistance of the body results in a slow-moving swimmer, and therefore small perturbations to the shape of filament due to swimming-induced additional drag. In that case, everywhere along the filament bending forces balance the drag due to the actuation and the bending amplitude, $y$, is on the order of $\epsilon \ell_\omega$.
The swimming velocity, $U$, is then found by balancing the drag on the swimmer body by the typical filament propulsive force  \cite{WigginsGoldstein,yu06}  
\begin{equation}
\mu a  U \sim A \int_0^L \frac{\p y}{\p x} \frac{\p^4 y }{\p x^4}\,\d x \sim 
 A\left( \frac{y}{\ell_\omega^2}\right)^2\sim \epsilon^2 \xi \omega \ell_\omega^2,
\end{equation}
and therefore
\begin{equation}\label{largea}
 U \sim \epsilon^2 \omega \frac{\ell_\omega^2}{a}\frac{1}{\log (\ell_w/r)},
\end{equation}
where the (slow) logarithmic dependence arises from the drag coefficient, $\xi \sim \mu/\log(\ell_\omega/r)$, where $\mu$ is the fluid viscosity. 

In the case of small body size, $a\ll \ell_\omega$, the expected scaling is more difficult to derive and requires a proper look at the local and global force balance for the swimmer. Let us denote by $V$  the transverse  velocity and by $\Omega$ the out-of-plane rotation rate of the swimmer, as measured as the center of mass of its body. Both $V$ and $\Omega$ average to zero over one actuation period but play a significant role nonetheless. Let us denote by $\ell$ the typical length scale along the filament where bending of the filament, of amplitude $y$,  is concentrated.
The  transverse force balance and out-of-plane torque balance  on the swimmer body lead to the scalings
\begin{equation}
\mu a (V +a \Omega)\sim A \frac{y}{\ell^3},\quad \mu a^2(V+a \Omega) \sim A \frac{y}{\ell^2},
\end{equation}
and therefore
\begin{equation}\label{VO}
V\sim a \Omega
,\quad
\Omega\sim \frac{Ay}{\mu}\left(\frac{1}{a^2\ell^3}+\frac{1}{a^3\ell^2}\right) \cdot
\end{equation}
Close to the hinge point, the drag due to the actuation is small and therefore the local force balance along the filament is between bending and drag due to the solid-body motion with velocity $V$, so that we expect
\begin{equation}
\xi V \sim A \frac{y}{\ell^4},  
\end{equation}
and therefore, given Eq.~\eqref{VO},
\begin{equation}
\frac{1}{\ell^2}\sim \frac{1}{\log (\ell_\omega/r)}\left(\frac{1}{a\ell}+\frac{1}{a^2}\right),
\end{equation}
which has the solution
\begin{equation}\label{ell}
\ell \sim a 
\left[\log \left(\frac{\ell_\omega}{r}\right)\right]^{1/2}\cdot
\end{equation}
The deformations of the filament are therefore concentrated in a (small) region of the size of the body (to within logarithmic terms). The bending amplitude $y$, is finally determined by local force balance along the filament beyond the small region of size $\ell$, where only the drag forces balance so that $ \Omega \sim \epsilon \omega$, and therefore, given Eqs.~\eqref{VO} and \eqref{ell}, we get
\begin{equation}
y\sim \epsilon \frac{a^5}{\ell_\omega^4}
\left[\log \left(\frac{\ell_\omega}{r}\right)\right]^2.
\end{equation}
In that case, the rotation rate exactly counterbalances the actuation and results in a straight unperturbed filament in the laboratory frame beyond the actuation region of size $\sim \ell$ near the actuation point. 
Finally, the swimming speed is found by balancing the drag on the filament by the propulsive force
\begin{equation}
\xi \ell_\omega U \sim A \int_0^L \frac{\p y}{\p x} \frac{\p^4 y }{\p x^4}\,\d x \sim A\epsilon \frac{y}{\ell^3},
\end{equation}
as the slope of the (almost straight) filament is set by the actuation ($\p y/\p x \sim \epsilon$) and the bending forces are non zero only in the region of size $\ell$ (with $\p^4 y/\p x^4 \sim y^4/\ell^4$). Consequently, we get the scaling
\begin{equation}\label{smalla}
U \sim \epsilon^2 \omega \frac{a^2}{\ell_\omega}\left[\log \left(\frac{\ell_\omega}{r}\right)\right]^{1/2}\cdot
\end{equation}
Note that this result implies that an elastic swimmer with no head cannot swim. The results of Eqs.~\eqref{largea} and \eqref{smalla} show therefore that the swimmer velocity is small for both large and small body size, and therefore an optimal body size should exist. This will be confirmed by calculations presented in \S\ref{S:optimal}.

\section{Solving for the kinematics of swimming}
\label{main}

We now turn to the analytical  calculations of the swimming kinematics of the elastic swimmer.  We first  present the assumptions used in this paper in \S\ref{S:assumptions}. The intrinsic formulation of the equations of motion is derived in  \S\ref{S:intrinsic} and simplified in \S\ref{S:smallslop} using the small-slope approximation. Using swimming kinematics defined in \S\ref{S:kinematics}, we can derive the free-swimming equations (\S\ref{S:freeswimming}), nondimensionalize the equations (\S\ref{S:nonD}) and solve for the shape of the oscillating filament together with the transverse velocities and rotation rates (\S\ref{S:shape}). The values of the axial velocity and rotation rates are then calculated in \S\ref{S:axial}
and the final expressions for the laboratory-frame kinematics are given in \S\ref{S:lab}.
Finally, the hydrodynamic efficiency of the swimmer is calculated in \S\ref{S:efficiency}.

\subsection{Assumptions}
\label{S:assumptions}
The calculations presented in this paper will be made under several simplifying assumptions.

First, the hydrodynamics is  simplified to the level of resistive-force theory \cite{gray55,keller76,lighthill76,brennen77}, a version of the equations of slender-body hydrodynamics \cite{tillett70,batchelor70,cox70,keller76-jfm,geer76,johnson80} where only the leading term in an expansion of hydrodynamic forces and moments in powers of $1/\log ({L/r})$ is conserved. In that case, the filament hydrodynamics is completely described by two drag coefficients, $\cpe$ and $\cpa$, relating linearly the  drag forces per unit length of the filament  to the local velocity relative to the fluid, for motion perpendicular and parallel to the filament respectively.  This widely used approximation is asymptotically valid in the limit of very slender filaments $L\gg r$, and results in theoretical predictions in quantitative agreement with propulsive force measurements  for actuated filaments \cite{yu06}. To simplify the analysis, we will also ignore hydrodynamic interactions between the swimmer body and the oscillating filament.

In the case of planar actuation, we then make the assumption in this paper that the amplitude of the actuation is small. More precisely, we will denote by $\epsilon$ the amplitude of the oscillation of the filament slope and will derive the swimming kinematics in the limit where $\epsilon\ll 1$. This has the advantage that the entire problem can be solved analytically and therefore shows directly the variations of swimming speeds and rotation rates with the various parameters of the problem. As was shown in the previous experimental study of the propulsive force mechanism by Yu {\it et al.} \cite{yu06}, the small slope approximation gives results which agree quantitatively with numerical and experimental results even for large slopes, and therefore we expect the results of this paper to remain valid up to  $\epsilon \sim 1$.

In the case where the actuation of the elastic filament is three-dimensional, we will furthermore neglect twist strains which will  be generated along the filament. Obviously, the precise nature of the twist strain density in the filament  depends on the exact actuation mechanism at the base of the filament. However, it is possible to show that such twist strains do not influence the kinematics of the swimmers in the limit which is of interest to us. Indeed, as a difference with bending amplitude which hyperdiffuses as described in Eq.~\eqref{hyperdiff}, twist strains satisfy a diffusion equation, and the intrinsic twisting-hydrodynamics length scale is $L_\omega = (C/\omega \xi_r)^{1/2}$, where $C$ is the twisting modulus  and $\xi_r$ is the rotational drag coefficient for the filament, that is $\xi_r=4\pi \mu r^2$  \cite{wolgemuth00}. Since the relevant filament length for effective swimming is $L\sim \ell_\omega$, and since for most solids $A\sim C$, we have $L/L_\omega\sim \ell_\omega/L_\omega \sim ({r}/{\ell_\omega} ) (\log (\ell_\omega/r))^{1/2}$. As the filament is slender, the typical twist penetration length scale is therefore much larger than the relevant filament length, $L/L_\omega \ll 1$, and as a consequence twist strains are always in diffusive equilibrium: They vary linearly between zero at the free-end of the filament and a constant at the base, which can be obtained by a local balance between viscous torque and twist. Now, with this balance, it is possible to show \cite{wolgemuth00} that the twist term in the equation of motion for the filament is smaller by a factor $r^2/\ell_\omega^2$ (within logarithmic terms) than the bending term, and therefore can safely be neglected. This also means that we can ignore in this paper possible  buckling (whirling) instabilities, which occur above a critical rotation frequency $\omega_c\sim A /\zeta_r L^2\sim A /\mu r^2 L^2 $ \cite{wolgemuth00}. Indeed, in the case where $L\sim \ell_\omega$, we get  $\omega_c/\omega \sim \ell_\omega^2/(r^2 \log(\ell_\omega/r)) \gg 1$, and therefore buckling instabilities appear for much larger values of the typical actuation frequency and can be neglected.
In that case, we will also ignore local rotational drag along the long filament \cite{powers02}.

\subsection{Intrinsic formulation and equations}
\label{S:intrinsic}

The elastic energy of the flexible filament as a function of its confirmation is given by
\begin{equation}\label{elasticenergy}
{\cal E} = \frac{1}{2} \int_0^L \left[ A \left(\frac{\p ^2 \r}{\p s^2}\right)^2  + \sigma \left( \frac{\p \r}{\p s} \right)^2  \right]\d s
\end{equation}
where $A$ is the bending modulus, $\sigma(s)$ the Lagrange multiplier (tension) which enforces local inextensibility and $s$ the curvilinear coordinate along the filament ($0\leq s \leq L$). As discussed above, twist strains are not included in this paper and therefore do not appear in Eq.~\eqref{elasticenergy}. In the case of filament of circular cross-section with radius $r$, we have $A=\pi r^4 E/4$, where $E$ is the Young's modulus of the material composing the filament.
Assuming resistive force theory for the fluids forces (see above), the intrinsic formulation for the filament elasto-hydrodynamics is obtained by calculus of variation from Eq.~\eqref{elasticenergy} and is given by
\begin{equation}
\label{equi}
[ \cpa \t \t + \cpe ({\bf 1}-\t \t ) ]\cdot \u = -A \frac{\p ^4 \r}{\p s^4} + \frac{\p }{\p s}\left( \sigma \frac{\p \r}{\p s}\right),
\end{equation}
where $\u$ is the local instantaneous velocity along the filament. The equation for the Lagrange multiplier $\sigma$, which enforces inextensibility $ (\r_s\cdot \r_s) _t= 0$ is given in an implicit form by
\begin{equation}\label{constraint}
{\bf t}\cdot \frac{\p {\bf u} }{\p s}=0.
\end{equation}
These two equations have boundary conditions which are given by
\begin{subeqnarray}
{\bf F}_{{\rm ext}} & = & A \frac{\p ^3\r}{\p s^3} - \sigma  \frac{\p \r}{\p s}, \quad {\bf t}\times \left[{\bf T}_{{\rm ext}} \times {\bf t} +A \frac{\p ^2\r}{\p s^2}\right]=0,\quad {\rm at }\quad  s=0, \\
{\bf F}_{{\rm ext}} & = & -A \frac{\p ^3\r}{\p s^3} + \sigma  \frac{\p \r}{\p s}, \quad {\bf t}\times\left[ {\bf T}_{{\rm ext}}\times {\bf t} -  A \frac{\p ^2\r}{\p s^2}\right]=0,\quad {\rm at }\quad  s=L,
\end{subeqnarray}
where ${\bf F}_{{\rm ext}}$ and ${\bf T}_{{\rm ext}}$ are externally applied forces and torques at the ends of the filament and ${\bf t} = {\bf r}_s$.

\subsection{Small-slope approximation}
\label{S:smallslop}
Let us consider a cartesian coordinate system $\{{\bf e}_x,{\bf e}_y,{\bf e}_z\}$ moving with the swimmer and located at the base of the elastic filament such that ${\bf e}_x$ is directed in the mean direction of the filament (see Fig.~\ref{mainfig}). The actuation oscillates therefore around the ${\bf e}_x$ axis.
The laboratory coordinate system is defined as $\{{\bf e}_1,{\bf e}_2,{\bf e}_3\}$ and is chosen such that the average swimming occurs in the ${\bf e}_1$ direction. The filament positions is described by the functions $y(x,t)$ and $z(x,t)$.
We assume that the filament shape is  slowly varying, that is 
\begin{equation}\label{approx}
\bigg\vert \frac{\p y}{\p x} \bigg\vert \ll 1,\quad
\bigg\vert\frac{\p z}{\p x} \bigg\vert \ll 1,
\end{equation}
 so that a point on the filament is defined as $\r \approx x{\bf e}_x + \rp$, where $\rp = y(x,t){\bf e}_y + z(x,t){\bf e}_z$, and we also have $s\approx x$.
In that case,  Eqs.~\eqref{equi} and \eqref{constraint}  become, when written in the frame moving with the swimmer,
\begin{subeqnarray}
\label{smallslope_general}
\cpa u_x + (\cpa-\cpe) \dyx u_y + (\cpa-\cpe)\dzx u_z  & = &  \frac{\p \sigma}{\p x} \slabel{fx},\\
(\cpa-\cpe) \dyx u_x + \cpe u_y + (\cpa-\cpe)\dyx\dzx u_z &  = & -A\frac{\p^4 y}{\p x^4}+ \frac{\p }{\p x}\left(\sigma\frac{\p y}{\p x}\right) \slabel{fy},\\
(\cpa-\cpe) \dzx u_x  + (\cpa-\cpe)\dyx\dzx u_y + \cpe u_z &  = & -A\frac{\p^4 z}{\p x^4}+ \frac{\p }{\p x}\left(\sigma\frac{\p z}{\p x}\right) \slabel{fz},\\
\frac{\p ^2 \sigma }{\p x^2}+A\left(\frac{\cpa}{\cpe}-1\right)
\frac{\p ^2 \rp}{\p x^2}\cdot \frac{\p ^4 \rp}{\p x^4} & =&  A \frac{\p  \rp}{\p x}\cdot \frac{\p ^5 \rp}{\p x^5},
\slabel{tension}
\end{subeqnarray}
where, $\{u_x,u_y,u_z\}$ are the local velocity components of the filament. 
The leading-order boundary conditions at either end of the filament are given, for all times, by
\begin{subeqnarray}
{\bf F}_{{\rm ext}} & = & A\frac{\p ^3 \rp}{\p x ^3}-\sigma\frac{\p \rp}{\p x}-\sigma  {\bf e}_x,\quad T_{{\rm ext},\, y}  = A \frac{\p^2 z}{\p x^2},\quad T_{{\rm ext},\, z}  = -A \frac{\p^2 y}{\p x^2} ,\quad {\rm at }\quad  s=0, \slabel{s0}\\
{\bf F}_{{\rm ext}} & = & -A\frac{\p ^3 \rp}{\p x ^3}{\bf e}_y +\sigma\frac{\p \rp}{\p x} +\sigma  {\bf e}_x,\quad T_{{\rm ext},\, y}  = -A \frac{\p^2 z}{\p x^2},\quad T_{{\rm ext},\, z}  = A \frac{\p^2 y}{\p x^2} ,\quad {\rm at }\quad  s=L. \slabel{sL}
\end{subeqnarray}
The most general periodic actuation at the frequency $\omega$ is 
\begin{subeqnarray}\label{origin}
y(0,t)=0, \quad \dyx(0,t)  =  \epsilon \cos \omega t,\\
z(0,t)=0, \quad \dzx(0,t)  =  \delta \epsilon \sin \omega t.
\end{subeqnarray}
where $0\leq \delta \leq 1$ measures the extent of the three-dimensionality of the actuation.
In that case,  $\delta=0$ represents a purely planar actuation, whereas for $\delta=1$,  the actuating hinge sweeps a cone. 
With these notations, the small slope approximation of Eq.~\eqref{approx} is written $\epsilon \ll 1$.

\subsection{Swimming kinematics}
\label{S:kinematics}
Let us denote by ${\bf U}(t)$ the instantaneous velocity of the swimmer body and $\boldsymbol{\Omega}(t)$ its instantaneous rotation rate around the hinge point \footnote{We find it more convenient in this paper to measure rotation rates and torque at the actuation point instead of the center of mass of the swimmer body. Obviously, this does not impact our final results.}. The motion of the filament is then described as a superposition of a solid body translation at velocity ${\bf U}$, a solid body rotation with rotation rate  $\boldsymbol{\Omega}$, and a relative motion due to the oscillations of the filament and described by the functions $\p y / \p t$ and $\p z / \p t$. As a consequence, the local velocity components along the filament are given by
\begin{subeqnarray}
u_x & = & U_x + z\Oy-y\Oz,\\
u_y & = & U_y+\dyt+x\Oz-z\Ox,\\
u_z & = & U_z+\dzt+y\Ox-x\Oy \cdot
\end{subeqnarray}

\subsection{Free-swimming assumption}
\label{S:freeswimming}

We consider a free-swimmer in Stokes flow, so the total force and torque on the swimmer must vanish. This means that  the forces and torques in Eqs.~\eqref{s0} have to balance the hydrodynamic forces and torques on the body of the swimmer, whereas the forces and torques in Eqs.~\eqref{sL} must vanish. In this paper the swimmer will be assumed to have an  axisymmetric body around the $x$ axis. This allows the mathematical formulation to remain manageable and the generalization to more complex body shapes is straightforward. In the center of mass of the swimmer body, the resistance matrices for fluid forces and torques acting on the body are written, by symmetry, as
\begin{equation}
\tilde{{\bf F}}_{{\rm ext}} = -
\left(
\begin{array}{ccc}
 \RFUxx &  0 & 0   \\
0  &   \RFUyy&   0\\
 0 & 0  &  \RFUzz 
\end{array}
\right)\cdot\tilde {\bf U}
,\quad
\tilde{{\bf T}}_{{\rm ext}} = -
\left(
\begin{array}{ccc}
\RLOxx &  0 & 0   \\
0  &   \RLOyy&   0\\
 0 & 0  &  \RLOzz 
\end{array}
\right)\cdot \tilde {\boldsymbol{\Omega}},
\end{equation}
where everything is measured relative to the center of mass of the body.
In the frame of reference located at the base of the filament (see Fig.~\ref{mainfig}), these relations are now  written as
\begin{eqnarray}
{\bf F}_{{\rm ext}} & = & -
\left(
\begin{array}{ccc}
 \RFUxx &  0 & 0   \\
0  &  \RFUyy&   0\\
 0 & 0  & \RFUzz 
\end{array}
\right)\cdot{\bf U}
+\left(
\begin{array}{ccc}
0 &  0 & 0   \\
0  & 0&   a\RFUyy\\
 0 & -a\RFUzz  & 0
\end{array}
\right)\cdot \boldsymbol{\Omega}, \\
{\bf T}_{{\rm ext}}  & = & 
\left(
\begin{array}{ccc}
0 &  0 & 0   \\
0 & 0 &  -a\RFUzz\\
 0 &  a\RFUyy  & 0 
\end{array}
\right)\cdot{\bf U}
-\left(
\begin{array}{ccc}
\RLOxx &  0 & 0   \\
0  &  \RLOyy + a^2\RFUzz&   0\\
 0 & 0  &  \RLOzz +a^2\RFUyy
\end{array}
\right)\cdot \boldsymbol{\Omega}, 
\end{eqnarray}
where $a$ is the distance between the center of mass of the swimmer body and the actuation point (see Fig.~\ref{mainfig}) and $\boldsymbol{\Omega}$ the rotation rate around the actuation point.
Using these notations, the force-free and torque-free conditions at $s=0$, Eq.~\eqref{s0}, are written as
\begin{subeqnarray}
\RFUxx U_x(t) & = & \sigma(0,t), \slabel{Fx}\\
\RFUyy U_y (t) -a \RFUyy   \Oz (t) & = & \di	-A\frac{\p^3 y}{\p x^3}(0,t)+\sigma\frac{\p y}{\p x}(0,t), \slabel{Fy}\\
-a \RFUyy U_y(t) +(\RLOzz+a^2\RFUyy )  \Oz (t) & = & \phantom{-}\di	A\frac{\p^2 y}{\p x^2}(0,t),\slabel{Lz}\\
\RFUzz U_z  (t)+a\RFUzz  \Oy (t) & = & \di	-A\frac{\p^3 z}{\p x^3}(0,t)+\sigma\frac{\p z}{\p x}(0,t),\slabel{Fz}\\
a\RFUzz U_z  (t)+(\RLOyy + a^2\RFUzz) \Oy (t) & = & \di	-A\frac{\p^2 z}{\p x^2}(0,t), \slabel{Ly}
\end{subeqnarray}
while the conditions at $s=L$ (Eq.~\ref{sL}) become
\begin{equation}\label{zeros}
\frac{\p^2  y}{\p x^2}(L,t) =\frac{\p^2 z }{\p x^2} (L,t)=\frac{\p^3 y }{\p x^3} (L,t)=\frac{\p^3 z }{\p x^3} (L,t) =\sigma(L,t) =0.
\end{equation}
The system of partial differential equations is closed by writing down the overall torque balance in the $x$-direction, leading to an equation for $\Omega_x$ 
\begin{equation}\label{Lx}
\RLOxx \Ox = \int_0^L \left[A \left( y\frac{\p ^4 z}{\p x^4} - z\frac{\p ^4 y}{\p x^4} \right) + z\frac{\p}{\p x}\left(\sigma \frac{\p y}{\p x}\right)-y\frac{\p}{\p x}\left(\sigma \frac{\p z}{\p x}\right)\right] \,\d x 
= A \left[ \frac{\p y}{\p x}\frac{\p^2 z}{\p x^2}- \frac{\p z}{\p x}\frac{\p^2 y}{\p x^2}\right]_{x=0}
\end{equation}
where we have used integration by parts and conditions \eqref{origin} and \eqref{zeros}.
We have now as many equations as we have unknowns. There are nine unknowns, $\{U_x,U_y,U_z,\Ox,\Oy,\Oz,\sigma,y,z\}$ and nine equations (Eqs.~\ref{fx}, \ref{fy}, \ref{fz}, \ref{tension}, \ref{Fy}, \ref{Lz}, \ref{Fz}, \ref{Ly}, \ref{Lx}). The equation for $\sigma$ is second order, and is accompanied by two boundary conditions (Eqs.~\ref{Fx} and \ref{zeros}). The equations for $y$ and $z$ are fourth-order and are accompanied by four boundary conditions each
(Eqs.~\ref{origin} and \ref{zeros}).

\subsection{Nondimensionalization and simplifications}
\label{S:nonD}

We now nondimensionalize the equations of motion. We scale  lengths by the intrinsic length scale $\ell_\omega$, time by $\omega^{-1}$, rotation rates by $\omega$, velocities by $\omega \ell_\omega$, resistivities by $\cpe \ell_\omega^n$ ($n=1$ or $3$ depending if it is a force-velocity or a torque-rotation rate resistivity), forces by $\cpe \ell_\omega^2 \omega $ and torques by $\cpe \ell_\omega^3 \omega $. 

Furthermore, we can use the fact that $\epsilon$ is small to simplify the dimensionless equations further. Let us evaluate the leading order power of $\epsilon$ for each of our nine unknowns. From Eq.~\eqref{origin} we see that $y\sim z\sim \epsilon$. Consequently, from Eqs.~\eqref{Fy}, \eqref{Lz}, \eqref{Fz}, and  \eqref{Ly} we see that $U_y\sim U_z\sim\Oy\sim\Oz\sim\epsilon$. We then get from Eq.~\eqref{Lx} that $\Ox\sim\epsilon^2$ and Eq.~\eqref{tension} shows that $\sigma\sim\epsilon^2$, so that Eq.~\eqref{Fx} leads to $U_x\sim\epsilon^2$.  The magnitude of the axial swimming speed and rotation rate are therefore one order of magnitude smaller than the transverse velocities and rotation rates.
These scalings allow one to simplify the equations for $y$ and $z$ further, and we obtain, using the same symbols as for the dimensional variables for simplification, 
\begin{subeqnarray}\label{new}
 \dyt + U_y(t) + x\Oz(t) & = & -\frac{\p^4 y}{\p x^4} \slabel{newy},\\
 \dzt + U_z(t) - x\Oy(t) & = & -\frac{\p^4 z}{\p x^4} .
\end{subeqnarray}
Because of the $\epsilon$ scaling for the transverse problem is an order of magnitude larger than the scaling $\epsilon^2$ for the axial problem, we see that the axial unknowns ($U_x,\Ox$) have disappeared from Eq.~\eqref{new}. We can therefore solve these two  problems in two separate times. First, we solve Eq.~\eqref{new} for the filament shape $\{y,z\}$ and the transverse swimming kinematics $\{U_y,U_z,\Oy,\Oz\}$, with boundary conditions
\begin{subeqnarray}\label{bc_dimensionless}
y(0,t)=0, \quad \dyx(0,t)  =  \epsilon \cos t, \quad \ddyx(\L,t)=0,\quad  \dddyx(\L,t)=0, \slabel{bcy}\\
z(0,t)=0, \quad \dzx(0,t)  =  \delta \epsilon \sin t, \quad \ddzx(\L,t)=0,\quad  \dddzx(\L,t)=0,
\end{subeqnarray}
where $L$ now refers to the dimensionless length of the filament. The dimensionless resistance  equations  become, with all the symbols referring now to dimensionless variables,
\begin{subeqnarray}
\RFUyy U_y (t) -a \RFUyy   \Oz (t) & = & \di	-\frac{\p^3 y}{\p x^3}(0,t) \slabel{bc_newy},\\
-a \RFUyy U_y(t) +(\RLOzz+a^2\RFUyy )  \Oz (t) & = & \phantom{-}\di	\frac{\p^2 y}{\p x^2} (0,t), \\
\RFUzz U_z  (t)+a\RFUzz  \Oy (t) & = & \di	-\frac{\p^3 z}{\p x^3} (0,t),\\
a\RFUzz U_z  (t)+(\RLOyy + a^2\RFUzz) \Oy (t) & = & \di	-\frac{\p^2 z}{\p x^2} (0,t),
\end{subeqnarray}
which can be inverted to give
\begin{subeqnarray}\label{transversekinematics}
U_y (t) & = & - \left(\frac{\RLOzz+a^2\RFUyy}{\RFUyy\RLOzz}\right)\frac{\p^3 y}{\p x^3}(0,t) +\frac{ a }{\RLOzz}\frac{\p^2 y}{\p x^2}(0,t),\\
\Oz (t) & = &	- \frac{a}{\RLOzz}\frac{\p^3 y}{\p x^3}(0,t) +\frac{  1 }{\RLOzz}\frac{\p^2 y}{\p x^2}(0,t),	\\
U_z(t) & = &	-\left(\frac{\RLOyy + a^2\RFUzz}{\RFUzz\RLOyy}\right)\frac{\p^3 z}{\p x^3}(0,t)+\frac{a}{\RLOyy}\frac{\p^2 z}{\p x^2}(0,t),\\
\Oy (t) & = & \frac{a}{\RLOyy}\frac{\p^3 z}{\p x^3}(0,t)-\frac{1}{\RLOyy}\frac{\p^2 z}{\p x^2}(0,t).
\end{subeqnarray}
Once we have the solution for $\{y,z,U_y,U_z,\Oy,\Oz\}$, we can use Eqs.~\eqref{fx}, \eqref{Fx} and \eqref{Lx} to get $U_x$ and $\Ox$. 
Note that the dimensionless version of Eq.~\eqref{fx} is written as
\begin{equation}\label{fx_nodim}
 u_x + (1-\g)\left( \dyx u_y + \dzx u_z\right)   =\g   \frac{\p \sigma}{\p x}
\end{equation}
where $\g = \cpe/\cpa$. Integration of Eq.~\eqref{fx_nodim} along the filament,  using Eqs.~\eqref{Fx} and \eqref{new}, leads to the formula we will use for the axial swimming velocity,  $U_x$, as
\begin{equation}\label{Ux_nodim}
(\g\RFUxx + \L )U_x = \Oz\int_0^\L y\,\d x -\Oy\int_0^\L z\,\d x +
 (1-\g)\left[\frac{1}{2}\left(\frac{\p^2 y}{\p x^2}\right)^2 - \frac{\p y}{\p x} \frac{\p^3 y }{\p x^3} +  \frac{1}{2}\left(\frac{\p^2 z}{\p x^2}\right)^2 - \frac{\p z}{\p x} \frac{\p^3 z }{\p x^3} 
  \right]_{x=0}.
\end{equation}

It is  important to note at this point that the small-slope approximation has resulted in a partial simplification of the problem: The axial velocity and rotation rate of the swimmer being one order of magnitude smaller than the transverse velocities and rotation rates, the axial swimming kinematics is slaved to the transverse kinematics. The problem of determining the filament shape and the transverse swimming kinematics cannot, however, be simplified any further and both still have to be solved simultaneously.

\subsection{Solving the transverse problem: Filament shape and swimming kinematics}
\label{S:shape}

Let us now solve Eq.~\eqref{new}. Since the forcing is harmonic, we will solve these equations in Fourier space and write, for all variables, $A(x,t)=\Re\{\hat A (x) \exp(-it)\}$.
Using Eq.~\eqref{transversekinematics}, we have the relations
\begin{subeqnarray}\label{RHS}
U_y(t) + x\Oz(t) & = &-\left(\frac{\RLOzz+a(a+x)\RFUyy}{\RFUyy\RLOzz}\right)\frac{\p^3 y}{\p x^3}(0,t)  +  \di \frac{(a+x)  }{\RLOzz}\frac{\p^2 y}{\p x^2}(0,t), \\
U_z(t) - x\Oy(t) & = &-\left( \frac{\RLOyy + a(a+x)\RFUzz}{\RFUzz\RLOyy} \right) \frac{\p^3 z}{\p x^3}(0,t)+ \frac{(a+x)}{\RLOyy}\frac{\p^2 z}{\p x^2}(0,t).
\end{subeqnarray}
From Eqs.~\eqref{new} and \eqref{RHS} we see that the generic equation satisfied by both $y$ and $z$ is a hyperdiffusion equation forced by a first order polynomial whose coefficients depend on the boundary condition of the solution.  This integro-differential equation is a consequence of the problem-coupling discussed in the introduction of the paper and reflects the  nonlocal aspect of locomotion without inertia where, at all times, velocities and rotation rates adjust so that total forces and torques sum up to zero.

Let us denote by $\zeta(x;\beta, \sigma,\lambda,\mu,h)$, the solution to the differential equation
\begin{equation}
\label{newpde}
\left\{ -i  + \frac{\d^4 }{\d x^4}\right\} \zeta(x)
 =( \beta +\sigma x)  \frac{\d^3 \zeta}{\d x^3}(0) + (\lambda+ \mu x) \frac{\d^2 \zeta}{\d x^2}(0),
\end{equation}
with boundary conditions
\begin{equation}
\label{bc}
\zeta(0)=0,\quad \frac{\d \zeta }{\d x}(0)=1,\quad \frac{\d^2\zeta}{\d x^2}(h)=0,\quad  \frac{\d^3\zeta}{\d x^3}(h)=0.
\end{equation}
In that case, it is easy to see from Eqs.~\eqref{new} and \eqref{bc_dimensionless} that, if we define
\begin{equation} 
\zeta_\perp(x)=\zeta\left(x;\frac{1}{\RFUyy}+\frac{a^2}{\RLOzz},\frac{a}{\RLOzz},-\frac{a}{\RLOzz},-\frac{1}{\RLOzz},\L\right),
\end{equation}
then  we have
\begin{equation} 
\label{yz}
y (x,t)  =  \epsilon\Re\{e^{-it} \zeta_\perp(x)\},\quad 
z (x,t)  =  \delta \epsilon\Re\{ie^{-it} \zeta_\perp(x)\},
\end{equation}
and we note that $\hat z = i \delta \hat y$. 

The analytical solution to Eqs.~\eqref{newpde}-\eqref{bc} is given by
\begin{equation}\label{zeta}
\zeta(x)=
\sum_{n=0}^{3} A_n  e^{\alpha_n x} + Bx +C,
\end{equation}
where $\alpha_n=\exp(i(1+4n)\pi/8)$ ($0\leq n \leq 3$)
and
where the six constants satisfy the linear system
\begin{eqnarray}
 \left[
\begin{array}{cccccc}
1 & 1 & 1 & 1 & 0 & 1 \\
\alpha_0 & \alpha_1 & \alpha_2 & \alpha_3 & 1&0 \\
\alpha_0^2 e^{\alpha_0h} & \alpha_1^2 e^{\alpha_1h} & \alpha_2^2 e^{\alpha_2h} & \alpha_3^2 e^{\alpha_3h} & 0 & 0\\
\alpha_0^3 e^{\alpha_0h} & \alpha_1^3 e^{\alpha_1h} & \alpha_2^3 e^{\alpha_2h} & \alpha_3^3 e^{\alpha_3h} & 0 & 0\\
\alpha_0^2 (\lambda+\beta \alpha_0) & \alpha_1^2 (\lambda+\beta \alpha_1) & \alpha_2^2 (\lambda+\beta \alpha_2) & \alpha_3^2 (\lambda+\beta \alpha_3) & 0 & i \\
\alpha_0^2 (\mu + \sigma\alpha_0) & \alpha_1^2 (\mu + \sigma\alpha_1) & \alpha_2^2 (\mu + \sigma\alpha_2) & \alpha_3^2 (\mu + \sigma\alpha_3) & i & 0
 \end{array}
\right] 
\cdot
\left[
\begin{array}{l}
A_0\\
A_1\\
A_2\\
A_3\\
B\\
C
\end{array}
\right]=
\left[
\begin{array}{r}
0\\
1\\
0\\
0\\
0\\
0
\end{array}
\right]\cdot
\end{eqnarray}
With  this solution known, we get the transverse velocities and rotation rates
\begin{subeqnarray}\label{UyOz}
\hat U_y  & = & - \epsilon\left(\frac{\RLOzz+a^2\RFUyy}{\RFUyy\RLOzz}\right) \zeta_\perp'''(0) +\frac{ a\epsilon }{\RLOzz} \zeta_\perp''(0),\\
\hat \Omega_z  & = &	- \frac{a\epsilon}{\RLOzz} \zeta_\perp'''(0) +\frac{ \epsilon }{\RLOzz} \zeta_\perp''(0),	\\
\hat U_z & = &	i\delta \hat U_y,\\
\hat \Omega_y & = & -i\delta \hat \Omega_z.
\end{subeqnarray}

This completes the solution of the transverse problem, and we now have the expressions for both the filament shape (Eq.~\ref{yz}) and the transverse swimming kinematics (Eq.~\ref{UyOz}).

\subsection{Solving the axial problem: Swimming velocity and rotation rate}
\label{S:axial}

We can now solve the second problem, namely find the expression for the axial velocity and rotation rates. Using the Fourier-space notation defined above, it is then easy to see that the axial swimming velocity (Eq.~\ref{Ux_nodim}) has two components, $U_x=\langle U_x \rangle + U_x'$, where $\langle U_x \rangle$ is the steady component and $U_x'$ the zero-mean oscillatory component, which are  given by
\begin{subeqnarray}
\langle U_x \rangle  & = & \frac{\epsilon^2(1+\delta^2)}{2(\g\RFUxx + \L )}
\Re\left\{
\I \bar\Omega_z^*+(1-\g)\left(\frac{1}{2}|\zeta_\perp''(0)|^2-\zeta_\perp'(0)\zeta_\perp'''(0)^*\right)
\right\},\\
U_x' & = & \frac{\epsilon^2(1-\delta^2)}{2(\g\RFUxx + \L )}\Re\left\{
e^{-2it}\left[\I\bar\Omega_z
+(1-\g)\left(\frac{1}{2}\zeta_\perp''(0)^2-\zeta_\perp'(0)\zeta_\perp'''(0) \right)
\right]
\right\},
\end{subeqnarray}
where we have defined ${\cal I}=\int_0^\L \zeta_\perp(x)\,\d x$ and $\bar\Omega_z=\hat\Omega_z/\epsilon$ (which is of order one). Note that when $\delta=1$, the unsteady component is exactly equal to zero (although the trajectory is not - it is  a helix). This is because in this case, the shape of the filament is steady in the frame rotating with the hinge \cite{powers02}. 

The equations above can be simplified further by noting that the integration in space of  the Fourier-transform of Eq.~\eqref{newy}, using Eq.~\eqref{bc_newy} as  boundary condition,  leads to the value of $\I$ as given by
\begin{equation}\label{I}
\I = i\bar\Omega_z\left(a\RFUpe-\frac{\L^2}{2}\right)-i(\RFUpe+\L)\bar U_y,
\end{equation}
where  $\bar U_y=\hat U_y/\epsilon$. As a consequence, we get
\begin{equation}
\Re\left\{\I \bar\Omega_z^*\right\}= (\RFUpe+\L)\Im \left\{ \bar U_y \bar \Omega_z^*\right\} 
\end{equation}
and, given Eq.~\eqref{UyOz}, it is straightforward to show that
\begin{equation}
\Im \left\{ \bar U_y \bar \Omega_z^*\right\} = \frac{1}{\RFUpe\RLOpe}\Im\left\{\zeta_\perp''(0) \zeta_\perp'''(0)^*\right\},
\end{equation}
so that the expression for the axial swimming velocity is simplified  to
\begin{subeqnarray}
\slabel{Ux_middle}\langle U_x \rangle  & = & \frac{\epsilon^2(1+\delta^2)}{2(\g\RFUxx + \L )}
\left[
(1-\g)\Re
\left\{\frac{1}{2}|\zeta_\perp''(0)|^2-\zeta_\perp'(0)\zeta_\perp'''(0)^*\right\} + 
\left(\frac{\RFUpe+\L}{\RFUpe\RLOpe}\right)\Im\left\{\zeta_\perp''(0) \zeta_\perp'''(0)^*\right\}
\right],\\
U_x' & = & \frac{\epsilon^2(1-\delta^2)}{2(\g\RFUxx + \L )}\left[
(1-\g)\Re
\left\{e^{-2it}\left(\frac{1}{2}\zeta_\perp''(0)^2-\zeta_\perp'(0)\zeta_\perp'''(0) \right)
\right\}\right.\\
&& \quad \quad \quad \quad \quad \quad \quad \nonumber
\left.+(\RFUpe+\L)
\Im\left\{\bar U_y \bar \Omega _z e^{-2it} \right\}
+\left(\frac{\L^2}{2}-a\RFUpe \right)\Im\left\{\bar\Omega_z^2e^{-2it}\right\}
\right].
\end{subeqnarray}

Finally, the leading order  version of torque balance in the $x$ direction, Eq.~\eqref{Lx}, shows that $\Ox$ does not have any oscillating components (i.e. is steady) and is equal to
\begin{equation}
\Ox =\langle \Ox \rangle
=\frac{\delta\epsilon^2}{\RLOxx}\Im\left\{ \zeta_\perp' (0)\zeta_\perp''^*(0)\right\},
\end{equation}
which finishes the solution for the axial problem. We now have the expression of all three-components of swimming velocities and rotation rates in the frame moving with the swimmer body.

\subsection{Swimming kinematics in the laboratory frame}
\label{S:lab}

The final step in the calculation is to transform  the expressions we have for the swimming kinematics in the body-fixed frame to the laboratory-fixed frame.  The general calculation for this transformation is presented in Appendix \ref{appendix}. It is worth noting that the distinction between the two frames of references has rarely been discussed in the context of swimming micro-organisms but can have important consequences if not properly taken into account. In our case, using  the transformation given by Eqs.~\eqref{transform1} and \eqref{transform2} as well as the swimming velocities obtained above, we find the final formulae for the leading-order swimming speeds in the laboratory frame as given by
\begin{subeqnarray}\label{final_formulae}
\slabel{meanU1}
\langle U_1\rangle & = &
\frac{\epsilon^2(1+\delta^2)}{2(\g\RFUxx + \L )}
\left[
(1-\g)\Re
\left\{\frac{1}{2}|\zeta_\perp''(0)|^2-\zeta_\perp'(0)\zeta_\perp'''(0)^*\right\}
+\left(\frac{\RFUpe - \g \RFUpa}{\RFUpe\RLOpe} \right)\Im\left\{ \zeta_\perp''(0) \zeta_\perp'''(0)^*\right\}
\right],
\\
\langle U_2 \rangle & = & 0,\\
\langle U_3 \rangle & = & 0,\\
U_1' & = & \frac{\epsilon^2(1-\delta^2)}{2(\g\RFUxx + \L )}\left[
(1-\g)\Re
\left\{e^{-2it}\left(\frac{1}{2}\zeta_\perp''(0)^2-\zeta_\perp'(0)\zeta_\perp'''(0) \right)
\right\}\right.\\
&& \quad \quad \quad \quad \quad \quad \quad
\left.+(\RFUpe+\g\RFUpa+2\L)\Im\left\{\bar U_y \bar \Omega _z e^{-2it}\right\}
+\left(\frac{\L^2}{2}-a\RFUpe \right)\Im\left\{\bar\Omega_z^2e^{-2it}\right\}
\right], \nonumber 
\\
U_2' & = &\epsilon\left[  \cos(\Ox t) \Re \left\{e^{-it} \bar U_y
\right\} +\delta \sin(\Ox t) \Im \left\{e^{-it} \bar U_y
\right\}\right], \slabel{U2} \\
U_3 & = & 
\epsilon\left[  \sin(\Ox t) \Re \left\{e^{-it} \bar U_y
\right\} -\delta \cos(\Ox t)
\Im \left\{e^{-it} \bar U_y
\right\}\right].\slabel{U3}
\end{subeqnarray}
The expressions given in Eq.~\eqref{final_formulae} are the main results of this paper.
Note that when the drag is isotropic ($\g=1$, $\RFUpa=\RFUpe$), the mean swimming velocity is exactly equal to zero. Drag anisotropy is therefore crucial for locomotion without inertia \cite{becker03}. Note also that in Eqs.~\eqref{U2} and \eqref{U3}, time appears as a parameter as a result of the separation of time scales in the limit of small $\epsilon$ (see discussion in Appendix \ref{appendix}).

\subsection{Hydrodynamic efficiency}
\label{S:efficiency}

We define in this paper the efficiency of the motion, ${\cal E}$,  as the ratio of  useful work (defined as the work necessary to move the entire swimmer at the steady speed $\langle U_1 \rangle $) by the total work done by the swimmer, 
\begin{equation}
{\cal E}=\frac{\langle F_1\rangle  \langle U_1 \rangle}{\di \left \langle \int {\bf f}\cdot {\bf u}\,\d x\right\rangle},
\end{equation}
which becomes in dimensionless variables and at leading order in $\epsilon^2$
\begin{equation}
{\cal E}=\frac{\di \left(\RFUxx + \frac{1}{\g}\L\right)\langle U_1 \rangle^2}
{\left\langle
{\bf U}\cdot {\bf F} + {\bf \Omega}\cdot{\bf L}+
 \di \int_0^\L 
\left[ \left( \frac{\p^4 y}{\p x^4}\right)^2
+ \left( \frac{\p^4 z}{\p x^4}\right)^2\right]\,\d x
 \right\rangle }.
\end{equation}
The first term in the denominator is given by
\begin{equation}
 {\bf U}\cdot {\bf F} + {\bf \Omega}\cdot{\bf L}=  \RFUyy (U_y-a\Omega_z)^2 + \RFUzz (U_z+a\Omega_y) ^2 + \RLOyy \Omega_y^2 + \RLOzz \Omega_z^2.
\end{equation}
while the second term  can be evaluated in Fourier space and we obtain the efficiency as given by
\begin{equation}\label{finalEff}
{\cal E}=\frac{2\di \left(\RFUxx + \frac{1}{\g}\L\right)\langle U_1 \rangle^2}
{\epsilon^2(1+\delta^2)\left[
  \RFUpe |\bar U_y - a \bar\Omega_z|^2+\RLOpe |\bar\Omega_z|^2+ \di \int_0^\L |\zeta_\perp''''(x)|^2
\,\d x\right]
  },
\end{equation}
with $\langle U_1 \rangle $ given by Eq.~\eqref{meanU1}. Note that since  $\langle U_1 \rangle $ scales linearly with $(1+\delta^2)$ and appears squared in  Eq.~\eqref{finalEff}, we obtain the result that a three-dimensional actuation ($\delta\neq 0$) is always more efficient than a planar one. 

\subsection{Asymptotic limit of long filament}
With our analytical formulae, we can now derive the swimming velocity in the limit of a long filament
$L \gg 1$ (that is, $L\gg \ell_\omega$ in dimensional variables). Although we expect the swimming velocity to decrease to zero in this case, it is the  biologically relevant  limit for the motion of spermatozoa  \cite{brennen77,WigginsGoldstein}. Obviously, spermatozoa  use a different swimming mechanism as the one described in this paper, so our purpose is merely to be able to  compare  swimming performances.

In the limit of large body $a\gg 1$ ($a\gg\ell_\omega$ in dimensional variables), it is easy to see that the mean velocity, given by Eq.~\eqref{meanU1}, becomes
\begin{equation}
\langle U_1\rangle  = 
\epsilon^2 \frac{(1+\delta^2)(1-\g)}{(\g\RFUxx + \L )}\left(\frac{\sqrt{2}-1}{2\sqrt{2}}\right).
\end{equation}
In the limit of small body $a\ll 1$ ($a\ll\ell_\omega$ in dimensional variables), one needs  to write down Taylor expansions for each of the body resistivities in $a/\ell_\omega$, which is tedious but straightforward, and the mean swimming velocity is given by
\begin{equation}
\langle U_1\rangle  = 
\epsilon^2 \frac{(1+\delta^2)a\g }{L}\RFUpe (\RFUpa-\RFUpe)\sqrt{\frac{\sqrt{2}-1}{4\sqrt{2}}}.
\end{equation}


\begin{table}[t]
\begin{center}
\begin{tabular}{lccccccccc}
&&& Velocity (I)& \phantom{roo}& Velocity (II) & \phantom{roo}& Velocity (III)&  \phantom{roo}& Efficiency (IV) \\
\\
&\quad $\di \frac{a}{L}$\quad\quad\quad&\quad$\di \frac{L}{\ell_\omega}$\quad\quad\quad\quad& $\di \frac{|\langle U_1 \rangle|}{\epsilon^2}$ && $\di \frac{|\langle U_1 \rangle|}{\epsilon^2(2a+L)}$ && $\di \frac{|\langle U_1 \rangle|}{2\epsilon^2 a}$ && $\di \frac{{\cal E}}{\epsilon^2}$\\
\\
  & & &$(\times 10^{3})$& &$(\times 10^{3})$&& $(\times 10^{3})$&  &\\
\\
\hline
\\
Spherical body  & & && &&& &  &\\
\quad\quad Optimal I &0.30 &2.51 & {\bf 6.9} &   &  1.7& & 4.6 & & 0.122\%\\
\quad\quad Optimal II &0.29& 2.38 & 6.7 &&  {\bf 1.8}  &  &4.9& & 0.100\%\\
\quad\quad Optimal III & 0.18 & 2.83& 5.7  & &  1.5  &  &{\bf 5.7} && 0.079\% \\
\quad\quad Optimal IV\phantom{roo}& 0.37 &  2.70 & 6.1&   &1.3& & 3.1 &&{\bf 0.142\%}\\
\\
Elongated body\quad  \quad\quad & & &   & &&&& &\\
\quad\quad Optimal I &  0.62 &2.78 & {\bf 17.9} &&  2.9&& 5.2& & 0.418\%\\
\quad\quad Optimal II & 0.44 & 2.82 &16.7 &  &{\bf 3.1}& &6.7&    & 0.369\%\\
\quad\quad Optimal III & 0.24& 3.34 &12.6  && 2.6  && {\bf 7.9} && 0.336\%\\
\quad\quad Optimal IV & 0.49 & 3.20  & 16.9 && 2.6  && 5.4 &&{\bf 0.465}\%\\
\\
\hline
\end{tabular}
\end{center}
\caption{Geometrical and actuation characteristics of the optimal elastic swimmers (spherical body and elongated body of aspect ratio 500). Four different  quantities are optimized: The velocity in the laboratory frame (I), the velocity in  swimmer length per unit beat (II), the velocity in body length per unit beat (III), and the mechanical efficiency of the swimmer (IV).  Velocities and efficiencies are given for the planar swimmer ($\delta=0$) but the geometrical characteristics of the optimal swimmers are independent of the value of $\delta$. The swimmers are illustrated in Fig.~\ref{optimal:fig}.
\label{table1}} 
\end{table}

\begin{figure}[t]
\begin{center}
\includegraphics[width=.6\textwidth]{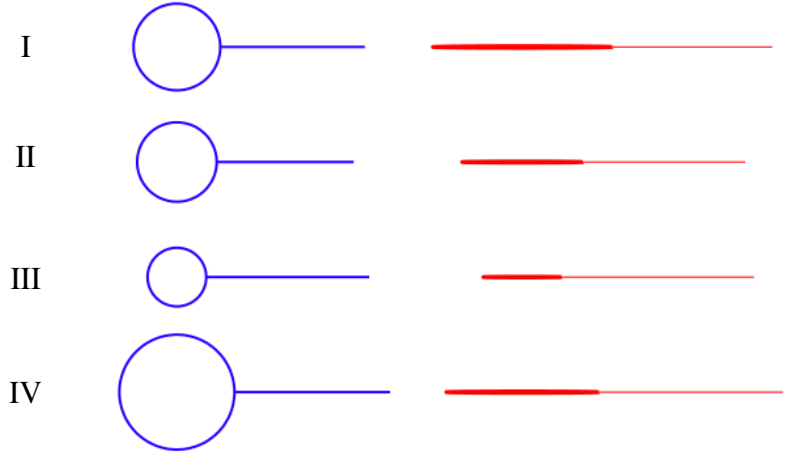} 
\caption{Optimal swimmers. Left (blue): Optimal swimmers with spherical body. Right (red): Optimal swimmers with elongated body.  (see caption of Table~\ref{table1} for description).}
\label{optimal:fig}
\end{center}
\end{figure}

\section{Performance and optimal design of the elastic swimmer}
\label{S:optimal}

As we have now analytical solutions for the complete swimming kinematics, we can study the performance of the elastic swimmer. 

\subsection{Optimal spherical swimmers}

As we have noted before, and as is confirmed by our analytical formulae,  swimming velocities and efficiencies decrease to zero for both large and small filament length and body size. As a consequence, optimal swimmers exist. The geometrical characteristics of these optimal swimmers are given in Table~\ref{table1} and represented schematically in Fig.~\ref{optimal:fig}.  For the calculations presented below, we have assumed the filament to be far from any boundaries and therefore the appropriate drag coefficients are given by 
\begin{equation}
\cpa  =  \frac{2\pi \mu}{\log (L/r) -1/2}, \quad \cpe  = \frac{4\pi \mu}{\log (L/r) +1/2},
\end{equation}
and we have chosen the aspect ratio of the filament to be $L/2r=500$ \footnote{The numerical  results depend very weakly on the value of the aspect ratio.}. We will furthermore assume the body to be a spheroid of revolution \cite{kimbook}.

We first determine the optimal swimmers with  spherical body   (Table~\ref{table1}  and Fig.~\ref{optimal:fig} left). There are a variety of ways to define the cost function to optimize, and we study four different optimality conditions. The first one is the swimming speed  $\langle U_1 \rangle$ (I). The second measure of performance is the swimming speed expressed in swimmer length per beat, that is $\langle U_1 \rangle/(2a+L)$ (II) . The third important velocity is the swimming velocity expressed in body length per unit beat, and is given by  $\langle U_1 \rangle/2a$ (III). Finally, we also consider the swimming efficiency as a measure of performance to optimize (IV). As shown in  Table~\ref{table1} and Fig.~\ref{optimal:fig}, the geometrical characteristics of the optimal swimmers are a strong function of the performance index which is chosen (but are independent of the value of $\delta$). In all cases, the optimal filament length $L$ is on the order of the intrinsic length scale $\ell_\omega$.

\begin{figure}[t]
\begin{center}
\includegraphics[width=.7\textwidth]{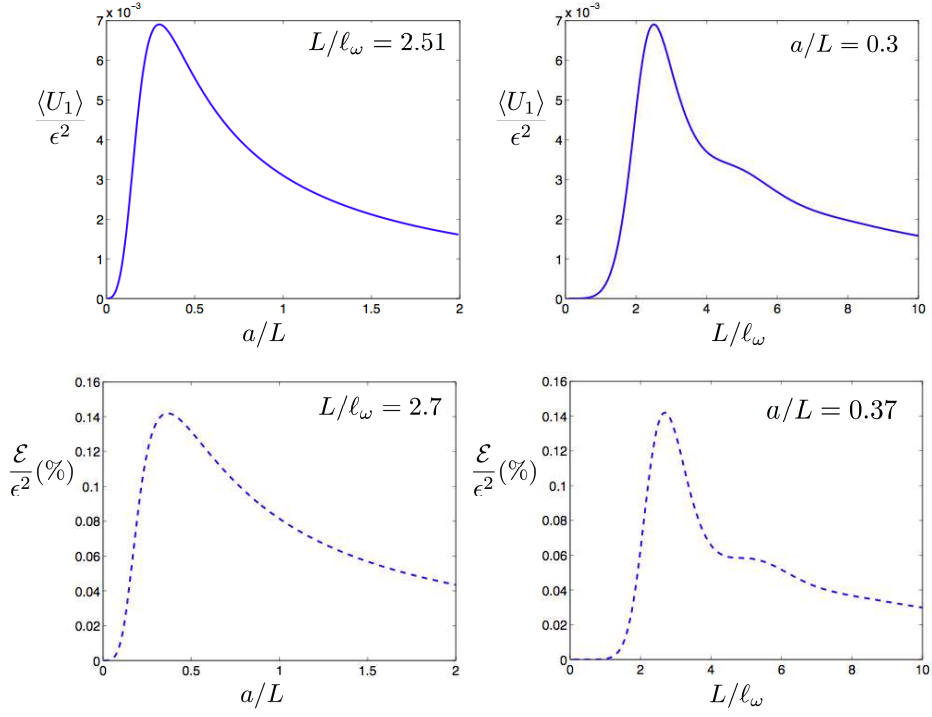}
\caption{Variation of the mean swimming velocity, $\langle U_1\rangle$, and the mechanical efficiency, ${\cal E}$, as a function of the  body to filament length ratio, $a/L$, and the dimensionless filament length, $L/\ell_\omega$, for the  optimal swimmer with spherical body I and IV (see Table~\ref{table1}). The values are displayed for the planar swimmer ($\delta = 0 $). Top: Variation of the swimming velocity for the fastest swimmer  ($a/L=0.3$, $L/\ell_\omega=2.51$); Bottom: Variation of the efficiency for the most efficient swimmer   ($a/L=0.37$, $L/\ell_\omega=2.7$).}
\label{variations}
\end{center}
\end{figure}

The variations of the swimming speed and efficiency of the optimal swimmer I (the fastest) and IV (the most efficient) with changing filament length and body size are displayed in Fig.~\ref{variations}. In particular, we see that the swimmer cannot move if either its filament or its body is too small. In the limit of large swimmer, the velocity also decreases to zero  as the inverse of the swimmer size.

It is interesting to note that, out of the three length scales which can be a priori chosen independently in designing an elastic swimmer - namely $a$, $L$, and $\ell_\omega$ - their relative magnitude  is fixed for the optimal swimmers (see Table~\ref{table1}) and therefore only one of them can be chosen arbitrarily. For a given performance index, once a specific body size or filament length or actuation frequency (through $\ell_\omega$) is chosen, everything else is fixed and there is only one possible optimal swimmer.

\subsection{Optimal swimmers}
We then study how the shape of the swimmer body influences the swimming performance and we find that a better performance is  always obtained for a long slender body of large aspect ratio in the swimming direction (the overall best is obtained in the limit of an infinite aspect ratio). We present in Table~\ref{table1} the performance of the optimal swimmer with an elongated body of aspect ratio 500, same aspect ratio as the filament. Both swimming speed and efficiency improve significantly by taking a slender swimming body. The optimal swimmers are displayed in Fig.~\ref{optimal:fig} (right). Again, the choice of a performance index has  consequences on the resulting optimal shape.

\subsection{Driving torque}
\begin{figure}[t]
\begin{center}
\includegraphics[width=.7\textwidth]{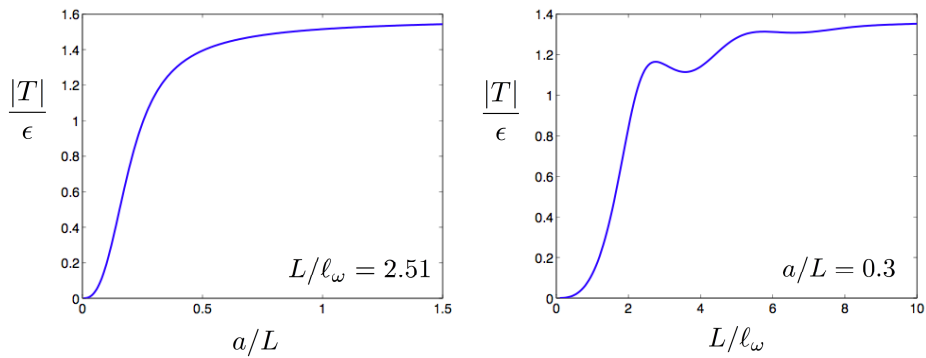}
\caption{Variation of the amplitude of the internal torque supplied by the actuating mechanism as a function of the body to filament length ratio, $a/L$, and the dimensionless filament length, $L/\ell_\omega$, for the  fastest swimmer with spherical body  ($a/L=0.3$, $L/\ell_\omega=2.51$, see Table~\ref{table1}) and for planar actuation ($\delta = 0 $).}
\label{torques}
\end{center}
\end{figure}

From a possible practical standpoint, it is important to quantify the internal torque necessary to drive the actuation at the given amplitude, $\epsilon$, and frequency, $\omega$. We  consider here the case of planar actuation. In that case, the oscillating torque is equal to the torque given in Eq.~\eqref{Lz} and therefore, in dimensionless form, the torque amplitude is given by $|T|=\epsilon |\zeta_\perp''(0)|$. The variation of the internal torque with the swimmer size is illustrated in Fig.~\ref{torques} in the case of the fastest swimmer with a spherical body (optimal swimmer I).

\subsection{Filament shape}

\begin{figure}[t]
\begin{center}
\includegraphics[width=.8\textwidth]{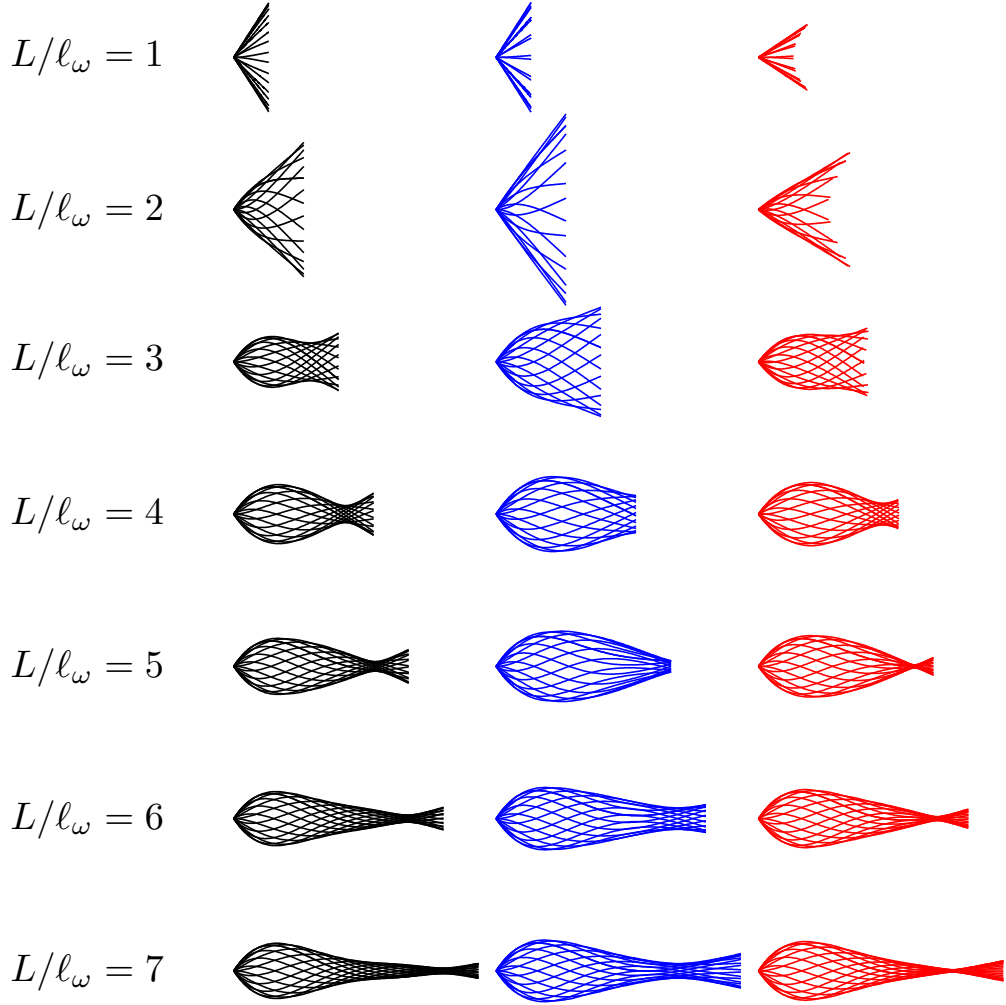}
\caption{Shapes of the elastic filament as a function of the dimensionless length $L/\ell_\omega$. Left (black): Shape of the filament in the case where the actuation point is fixed in space. In that case, the filament does not swim and its shape is that found theoretically by Wiggins and Goldstein \cite{WigginsGoldstein} and experimentally by Yu {\it et al.} \cite{yu06} . Center (blue): Shape of the filament in the case of the free-swimming as seen in the frame $\{{\bf e}_x,{\bf e}_y,{\bf e}_z\}$ translating and rotating with the swimmer body. The body size is $a/L=0.3$. Right (red): Shape of the filament for the same swimmer as seen in the laboratory frame $\{{\bf e}_1,{\bf e}_2,{\bf e}_3\}$. In that case, the instantaneous swimming velocities have been subtracted in order to be able to compare shapes.}
\label{shapes}
\end{center}
\end{figure}

Finally, we can use our analytical solution to characterize the shape of the flexible filament as the swimmers moves, and compare it to the shape obtained when the filament is actuated but does not swim \cite{WigginsGoldstein,yu06}. The results are shown in Fig.~\ref{shapes} for the planar swimmer with $a/L=0.3$. Since the swimmer body can always rotate to relieve some of the applied torque at the base of the filament, the curvature of the filament is smaller in the free-swimming case than it is in the case where the filament is not free to move (compare the  shapes on the left and at the center of Fig.~\ref{shapes}). We also note that the filaments in both cases display exponentially decaying amplitude, a feature which makes this problem intrinsically different from  eukaryotic flagellar propulsion where large-amplitude oscillations are present along the entire filament.
 
\section{Elastic swimming with more than one filament}
\label{S:many}
In this final section, we discuss possible improvements on the design of elastic swimmers. One  drawback of using a single filament as a propeller is the oscillatory nature of the swimming kinematics. As a result, a lot of effort goes into propelling the swimmer body in a direction which is different from the main swimming direction. In fact, in the limit of small actuation amplitude $\epsilon$, we have seen above that the mean swimming speed is of order $\epsilon^2$ whereas the transverse swimming speed (and rotation rate) are both of order $\epsilon$ and therefore one order of magnitude larger. We propose in this section to use more than one filament in order to have better control over the instantaneous swimming direction. For simplicity, we will ignore hydrodynamic interactions between the filaments in the analysis below.

\begin{figure}[t]
\begin{center}
\includegraphics[width=.8\textwidth]{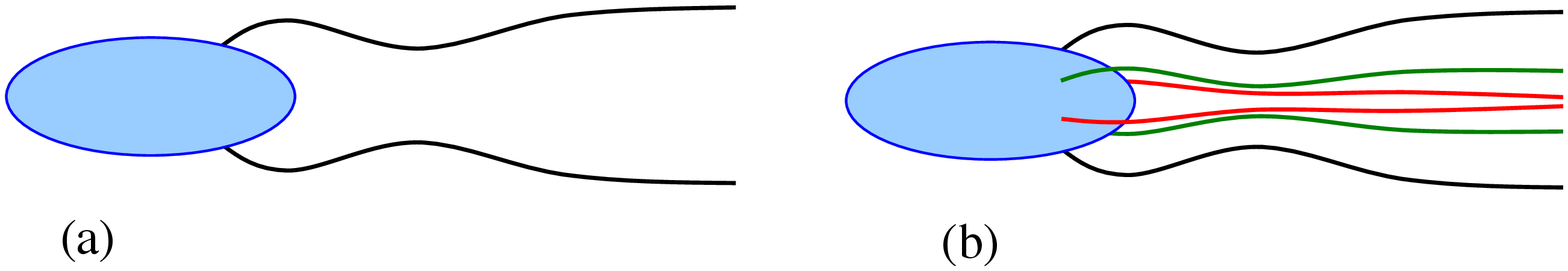}
\caption{Swimming with more than one elastic filament; (a) Swimming with two symmetric  filaments in planar motion  leads to unsteady straight swimming  (\S\ref{twoflagella}); (b) Swimming with three symmetric  filament pairs (so six filaments) in planar motion with each $2\pi/3$ phase difference leads to  steady straight swimming (\S\ref{sixflagella}).}
\label{fig_many}
\end{center}
\end{figure}

\subsection{Two filaments in planar motion}
\label{twoflagella}
If the swimmer has two filaments which are (a) positioned exactly symmetrically with respect to the axis of symmetry of the swimmer body and (b) are actuated with 180$^\circ$ out of phase, then all transverse forcing due to one filament will be exactly canceled by the second one, and this will result in  a straight (yet unsteady) swimming trajectory (see Fig.~\ref{fig_many}a). From a mathematical standpoint,  the dynamic in the $x$ and $y$ directions will be coupled for each filament, and we expect therefore a steady swimming of order $\sim\epsilon^2$ with oscillations of magnitude $\sim \epsilon$ along the same direction. In that case, the only nonzero component of the velocity is $U(t)$, along the ${\bf e}_1$ direction (the direction ${\bf e}_x$ still refers to the direction of each of the filaments) and there is no rotation rate. The equation for the shape of each filament becomes
\begin{equation}
 \dyt +\frac{\p^4 y}{\p x^4} = U \sin\theta   ,
\end{equation}
where $\theta$ is the angle between the average position of the filament base and the swimming direction. Force balance in the $x$ direction gives
\begin{equation}\label{U_straight}
(\g\RFUxx + 2 \L \cos^2\theta  )U(t) =
2\g\sin\theta  \left[	\frac{\p^3 y}{\p x^3}\right]_{x=0}+
 2 (1-\g) \cos\theta\left[\frac{1}{2}\left(\frac{\p^2 y}{\p x^2}\right)^2 - \frac{\p y}{\p x} \frac{\p^3 y }{\p x^3} 
  \right]_{x=0}.
\end{equation}
This is solved order by order for both $y(x,t)=\epsilon y_1 + \epsilon^2y_2+...$ and $U(t) = \epsilon U_1 + \epsilon^2 U_2+...$, and only $U_2$ has a non-zero time average. At leading order we find that $y_1$ is solution to the equation
\begin{equation}
\frac{\p y_1}{\p t}  + \frac{\p^4 y_1}{\p x^4}  = \frac{2\g \sin^2\theta}{\g\RFUxx + 2  \L\cos^2\theta}\left[\frac{\p^3 y_1}{\p x^3}\right]_{x=0},
\end{equation}
with boundary conditions 
\begin{equation}
y_1(0,t)=0, \quad \frac{\p y_1}{\p x}(0,t)  =  \cos t, \quad \frac{\p^2 y_1}{\p x^2}(\L,t)=0,\quad  \frac{\p^3 y_1}{\p x^3}(\L,t)=0.
\end{equation}
Consequently, $y_1$ is given by
\begin{equation}
y_1(x,t)= \Re\{e^{-it}\zeta_1(x)\},\quad \zeta_1(x)=\zeta\left(x;\frac{2\g \sin^2\theta}{\g\RFUxx + 2 \L \cos^2\theta},0,0,0,\L\right),
\end{equation}
and then the leading order component of the velocity can be calculated by
\begin{equation}
U_1(t)=\frac{2\g \sin\theta}{\g\RFUxx + 2 \L \cos^2\theta}  
\Re\{e^{-it}\zeta_1'''(0)\},
\end{equation}
which is purely oscillating (no mean component). We need to go to next order to calculate the mean component of the swimming. Since there is a component of the mean swimming velocity normal to the filament, each filament is therefore slightly asymmetric (the asymmetry, {\it i.e.} the mean value for $y_2$,  is of order $\epsilon^2$), a feature which is not present when we consider the case of a single filament. The second order term, $y_2$, satisfies the equation
\begin{equation}
\frac{\p y_2}{\p t}  + \frac{\p^4 y_2}{\p x^4}  = U_2(t)\sin\theta,
\end{equation}
with homogeneous boundary conditions. 
Since $U_2$ is expected to have both a mean value and an oscillating part, $U_2(t) = \langle U_2\rangle + U_2'$, we consider them separately. The oscillating part of $U_2$ will lead to an oscillating value for $y_2$ which will average out to zero in Eq.~\eqref{U_straight}, and therefore we do not need to solve for it. On the contrary the steady component of $U_2$ leads to a steady deflection of the filament, described by
\begin{equation}
 \frac{\d^4 \langle y_2 \rangle}{\d x^4}  = \langle U_2 \rangle \sin\theta,
\end{equation}
with homogeneous boundary conditions. This can be integrated to give
\begin{equation}\label{deflection}
\langle y_2 \rangle (x)=\langle U_2 \rangle  \sin\theta f(x),\quad f(x) = \left[ \frac{x^4}{24}-\frac{x^3\L}{6} + \frac{x^2\L^2}{4} \right],
\end{equation}
which can be then used to evaluate the mean swimming speed by averaging Eq.~\eqref{U_straight} at second order, and we find
\begin{equation}
(\g\RFUxx + 2 \L\cos^2\theta  + 2\g\L\sin^2\theta  )\langle U_2 \rangle = 
  (1-\g) \cos\theta 
     \left[\frac{1}{2}|\zeta_1''(0)|^2 -\Re\left\{ \zeta_1'(0) \zeta_1'''^*(0)\right\}
  \right].
\end{equation}
When  the swimmer does not have a body, $\RFUpa=0$, we obtain the ``elastic scallop'', flexible version of the two-arm swimmer discussed by Purcell and whose name is at the origin of the so-called scallop theorem \cite{purcell77}. Note that a two-filament swimmer with no head can swim, whereas a swimmer with a single filament cannot swim without a head.
In the case where $L\gg 1$, the function $\zeta_1$ can be simplified further \cite{WigginsGoldstein} and we find the average swimming velocity as given by 
\begin{equation}
\langle U \rangle =\frac{\sqrt{2}-1}{\sqrt{2}} \frac{ \epsilon^2 (1-\g) \cos\theta }{\g\RFUxx + 2 \L\cos^2\theta  + 2\g\L\sin^2\theta },
\end{equation}
with corrections exponentially small in $L/\ell_\omega$.

\subsection{Steady swimming: more than two filaments}
\label{sixflagella}

The setup with two filaments allowed us to get motion on a straight line. However, we obtained a steady  swimming speed of order  $\sim\epsilon^2$ with longitudinal oscillations of order  $\sim\epsilon$, which is not ideal. Here we ask the question, can we design a swimmer which moves on a straight line steadily? Since we look for straight motion, we will always look for pairs of filaments, located symmetrically with respect to the body symmetry axis (recall that the body is assumed to be axisymmetric), and mirror image to one another (see Fig.~\ref{fig_many}b).  We need  to find the minimum number of filament pairs together with the phase differences between each of them which is necessary to choose in order to achieve an overall steady propulsive force despite the unsteady propulsive forces (and shapes) of the individual filaments. When appropriately chosen, these phase differences will cancel out all of the $\sim\epsilon$ contributions from each individual filaments and should leave us with a steady $\sim\epsilon^2$ swimming speed.

The equation for each of the individual filaments in one pair ($1\leq n\leq N$, where $N$ is the number of pairs) is given by
\begin{equation}\label{yn}
 \frac{\p y_n}{\p t}+\frac{\p^4 y_n}{\p x^4} = U \sin\theta   ,
\end{equation}
with the boundary conditions, 
\begin{equation}
y_n(0,t)=0, \quad 
\frac{\p y_n}{\p x}(0,t)  =   \epsilon \Re\{c_n e^{-it}\}, \quad 
\frac{\p^2 y_n}{\p x^2}(\L,t)=0,\quad 
\frac{\p^3 y_n}{\p x^3}(\L,t)=0,
\end{equation}
where the set of complex constants $c_n$ is determining the phase difference in the actuation for each filament pair. Note that the filaments could be located anywhere around the swimmer body, but the assumption of  ignoring hydrodynamic interactions will be the most valid when they are the furthest apart, so we assume the filaments to be distributed with an angle $\pi/N$ apart from each other. Assuming a steady swimming speed, the solution to Eq.~\eqref{yn} is given formally by
\begin{equation}
\label{many}
y_n(x,t)=U\sin\theta f(x) +  \epsilon \Re \{c_n  e^{-it} g(x)\},
\end{equation}
where $f$ is the steady deflection above (Eq.~\ref{deflection}) and $g(x)$ is the Wiggins \& Goldstein shape (shape of the elastica when it is actuated at one end and does not swim), that is
\begin{equation}
g(x) = \zeta\left(x;0,0,0,0,\L\right). 
\end{equation}
The force balance in the swimming direction leads to the swimming speed as given by
\begin{equation}
\label{U_many}
(\g\RFUxx + 2N \cos^2\theta \L )U =
2\g\sin\theta \sum_{n=1}^N\left[	\frac{\p^3 y_n}{\p x^3}\right]_{x=0} +
 2 (1-\g) \cos\theta
  \sum_{n=1}^N
  \left[\frac{1}{2}\left(\frac{\p^2 y_n}{\p x^2}\right)^2 - \frac{\p y_n}{\p x} \frac{\p^3 y_n }{\p x^3} 
  \right]_{x=0}.
\end{equation}
Given that the filament shape appears in Eq.~\eqref{U_many} with linear and quadratic terms, it is straightforward to see that the smallest number of filament pairs necessary to achieve straight steady swimming is  such that
\begin{equation}
\sum_{n=1}^{N}c_n = 0, \quad \sum_{n=1}^{N}c_n^2 = 0.
\end{equation}
The minimum number of filament pairs is therefore $N=3$ (so 6 individual filaments) with $c_n=e^{2ni\pi/3}$ ($n=1,2,3$). The value of these constants show that if one pair of filaments displays base oscillations of the form $\cos t$ then the other two pairs need to oscillate  as $\cos (t\pm 2\pi/3)$. In that case, we obtain that the (steady) swimming speed is given by the quadratic equation
a quadratic equation
\begin{equation}
(\g\RFUxx + 6 \L\cos^2\theta  + 6\g \L\sin^2\theta  ) U =  \frac{3}{4}(1-\g) U^2\L^4\sin^2\theta \cos\theta 
+ 3\epsilon^2   (1-\g)\cos\theta \left[ \frac{1}{2}|g''(0)|^2-\Re\{g'(0)g'''^*(0)\}\right],
\end{equation}
and since we have assumed $\epsilon\ll 1$ we get the leading order solution given by
\begin{equation}\label{3pairs}
(\g\RFUxx + 6 \L \cos^2\theta + 6 \g\L \sin^2\theta  ) U =  
 3\epsilon^2   (1-\g)\cos\theta \left[ \frac{1}{2}|g''(0)|^2-\Re\{g'(0)g'''^*(0)\}\right].
 \end{equation}
Eq.~\eqref{3pairs} is the only instance where, up to geometric factors, the equation for the swimming speed is simply given by a balance between the viscous  drag on the filaments and the  propulsive forces from Wiggins \& Goldstein for all six filaments. In the limit $L\gg 1$, the approximate solution for the swimming velocity is now  given by
\begin{equation}
U = \frac{3(\sqrt{2}-1)}{\sqrt{2}}\frac{  \epsilon^2  (1-\g)\cos\theta}{\g\RFUxx + 6 \L \cos^2\theta + 6 \g\L \sin^2\theta}\cdot
\end{equation}


\section{Discussion}
\label{discussion}

We have presented in this paper an analytical treatment of the locomotion of an elastic swimmer in the limit of small amplitude actuation. This is arguably the simplest device which takes advantage of the coupling between drag and bending forces for locomotion purposes. Our study is different from and improves upon  previous work in many aspects: (a) Our analysis does not violate force balance nor torque balance and the constraints of free-swimming motion are fully enforced; (b) We include in our study the distinction between body-fixed and laboratory-fixed frames of references and calculate the swimming kinematics in both frames; (c) The coupling between the two problems - finding the shape of the filament and the swimming kinematics - is discussed for the first time and is solved in a self-consistent manner; (d) Our study produces analytical formulae; (e) We obtain the expected limits of vanishing swimming speed for small and large swimmer sizes; (f) We  characterize the geometry and performance of optimal swimmers; (g) The study of elastic swimming with more than one filament is presented, as it allows a better control on the swimming trajectories.

It is enlightening to compare the performance of the elastic swimmer with that of swimming micro-organisms - which of course use different actuation mechanisms. First, let us consider the bacterium {\it E. coli} \cite{bergbook}.  The helical flagella of {\it E. coli} are rotated at a frequency  $\omega\approx100$ Hz, resulting in a swimming velocity of about $U\approx30$ $\mu$m/s for a bacterium of size $L\approx 10$ $\mu$m. As a consequence, {\it E. coli} swims approximately at the speed of $U\approx 1/30$ body length per beat. As a difference, the flagellum of sea-urchin spermatozoon is actively oscillating  at a frequency $\omega\approx 40$ Hz, resulting in a velocity  $U \approx 200$ $\mu$m/s for an organism of size $L\approx 50$ $\mu$m \cite{brennen77}. The swimming speed in this case is therefore about  $U\approx 1/10$ body length per beat. How does that compare with our optimal swimmers? To answer this, we look at the results of Table~\ref{table1} for optimal swimmer II. Since we have nondimensionalized times by $\omega^{-1}$, we need to multiply the results by $2\pi$ to obtain velocities per unit frequency.
In the case $\epsilon \sim 1$, we find that the optimal swimmer with spherical body swims at about  $U\approx 1/90$ body length per beat whereas the optimal swimmer with slender body has a velocity  $U\approx 1/50$ body length per beat. The optimal elastic swimmers have therefore swimming performances which are comparable (although smaller by less than order of magnitude)  with that of real microorganisms. The most efficient elastic  swimmers  (typically $0.1-0.4\%$ efficiency) are also outperformed  by typical swimming microorganisms (usually $1-2\%$ of swimming efficiency). Both performances result from the exponentially decaying filament amplitude in the case of elastic swimmers, as compared with the large amplitude motion of real flagella.  

To conclude, we note that the results of our study could be improved upon in different ways. First, the treatment of the hydrodynamics of the filament using resistive-force theory is approximate and could be improved upon by  using slender-body theory -  most likely numerically. Hydrodynamic interactions between the filament and the swimmer body should also be included. Further improvement could be obtained by computing the swimming trajectories in the case of large-amplitude actuation and by including, in the case of three-dimensional actuation, the filament twist strains. Finally, the issue of thermal fluctuations (for the filament shape) and diffusion (for the swimmer position and orientation) should also be considered. Work in these directions is currently underway and will be reported in a future paper.

\section*{Acknowledgments}
We thank A. E. Hosoi and T. Yu for useful discussions. This work was supported in part by the Hock Tan Postdoctoral Fellowship in the Department of Mechanical Engineering at MIT.

\appendix
\section{Relationship between the body-fixed and the laboratory-fixed frames of reference}
\label{appendix}
In this appendix we derive the relationship between the swimming kinematics in the frame translating and rotating with the swimmer body and those in the laboratory frame of reference. The calculations are straightforward but have to be done correctly, and therefore are worth deriving. As in the main text of this paper, we denote by  $\{{\bf e}_x,{\bf e}_y,{\bf e}_z\}$ the cartesian coordinate system moving with the swimmer body and by  $\{{\bf e}_1,{\bf e}_2,{\bf e}_3\}$ the cartesian coordinate system in the laboratory frame of reference, with ${\bf e}_1$ being the  average swimming direction. In the limit of $\epsilon$ going to zero, we expect ${\bf e}_x$ to be almost equal to ${\bf e}_1$, and both ${\bf e}_y$ and ${\bf e}_z$ to be almost given by a solid body rotation around ${\bf e}_1$ at a constant rate, equal to $\Ox$. 

Let us  define the matrix 
\begin{equation}
E(t,\tau)=[{\bf e}_x \,\,{\bf e}_y\,\, {\bf e}_z]^T,
\end{equation}
then the equation  for the kinematics of the moving frame is given by
\begin{equation}\label{movingframe}
\frac{\d}{\d t}E = {\cal M}(t) \cdot E,
\end{equation}
with
\begin{equation}
{\cal M}(t)=
\left(
\begin{array}{ccc}
0 & \Oz(t) & -\Oy(t) \\
-\Oz(t)& 0& \Ox\\
\Oy(t)& -\Ox& 0
\end{array}
\right)\cdot
\end{equation}

The subtle issue which arises here is the appearance of a new (long) time scale $\sim 1/\Ox\sim 1/\epsilon^2$, time scale over which the $y$ and $z$ axis are expected to rotate by an angle $\pi/2$. Since the other time scale is $\sim 1$, then in order to obtain a solution for the dynamics of the body-attached frame for all times (both short and long time scales) in the limit of small $\epsilon$,  we have to use the method of multiple scales. 

We then define a new time $\tau=\epsilon^2 t$, so that $\d/\d t$ is formally replaced by $\p/\p t + \epsilon^2 \p/\p \tau$. 
In that case, Eq.~\eqref{movingframe}  becomes, in the multiple scale setting, 
\begin{equation}
\left(\frac{\p }{\p t} +\epsilon^2\frac{\p }{\p \tau} \right)E=(\epsilon M(t) + \epsilon^2 N) \cdot E,
\end{equation}
where we have used  the fact that both $\Oy$ and $\Oz$ are function of the short time scale and of order $\epsilon$, while $\Ox$ is constant and or order $\epsilon^2$. We then look for a solution as a perturbation expansion $E(t,\tau)=E_0 + \epsilon E_1 + \epsilon^2 E_2+...$, and obtain
\begin{subeqnarray}\label{orderbyorder}
\frac{\p E_0}{\p t} & = &0, \\
\frac{\p E_1}{\p t} & = & M(t)\cdot E_0,\\
\frac{\p E_2}{\p t} + \frac{\p E_0}{\p \tau}  & = & N\cdot E_0 + M(t)\cdot E_1,\\
\frac{\p E_3}{\p t} + \frac{\p E_1}{\p \tau}   & = & N\cdot E_1 + M(t)\cdot E_2.
\end{subeqnarray}
We then can solve the system given by Eq.~\eqref{orderbyorder} order by order, using the usual  multiple-scales trick that terms which would violate the perturbation expansion hypothesis have to be set to zero, and we obtain, written in the original time variable for the first two terms as
\begin{subeqnarray}\label{frame}
{\bf e}_x & =& {\bf e}_1 + \Re\left\{i\hat\Omega_z e^{-it} \right\} [ \cos(\Ox t) {\bf e}_2 + \sin(\Ox t) {\bf e}_3]- \Re\left\{i\hat\Omega_y e^{-it} \right\} [- \sin(\Ox t) {\bf e}_2 + \cos(\Ox t) {\bf e}_3],\\ 
{\bf e}_y & = &-\Re\left\{i\hat\Omega_z e^{-it} \right\}  {\bf e}_1 + \cos(\Ox t) {\bf e}_2 + \sin(\Ox t) {\bf e}_3, \\ 
{\bf e}_z & = &\Re\left\{i\hat\Omega_y e^{-it} \right\} {\bf e}_1 - \sin(\Ox t) {\bf e}_2 + \cos(\Ox t) {\bf e}_3,
 \end{subeqnarray}
where the notations introduced in the main part of this paper for the Fourier transforms have been used. We can now write down the relationship between velocities in the laboratory frame and velocities in the frame moving with the swimmer:
\begin{equation}
{\bf U}=U_x{\bf e}_x+U_y{\bf e}_y+U_z{\bf e}_z,
\end{equation}
which means, given Eq.~\eqref{frame}, that we have, at leading order for each component
\begin{subeqnarray}\label{transform1}
U_1 & = & U_x-U_y\Re\left\{i\hat\Omega_z e^{-it} \right\}+U_z\Re\left\{i\hat\Omega_y e^{-it} \right\},\\
U_2 & = & U_y \cos(\Ox t) - U_z \sin(\Ox t),\\
U_3 & = & U_y \sin (\Ox t)+ U_z \cos(\Ox t).
\end{subeqnarray}
Finally, denoting the averages on the short time scale by $\langle ...\rangle $, we get
\begin{subeqnarray}\label{transform2}
\langle U_1 \rangle  & = & \langle U_x\rangle +\frac{1}{2} \Im\{ \hat U_z\hat \Omega _y^*+\hat U_y^*\hat \Omega _z\},\\
U_1' & = & U_x' +\frac{1}{2}\Im \left\{ e^{-2it}(\hat U_y \hat \Omega _z - \hat U_z\hat\Omega_y ) \right\},
\end{subeqnarray}
and since both $U_y$ and $U_z$ average to zero, we obtain $\langle U_2\rangle = \langle U_3 \rangle = 0$.

\bibliographystyle{unsrt}
\bibliography{ElasticSwimming}

\begin{thebibliography}{10}

\bibitem{taylor51}
G.~I. Taylor.
\newblock Analysis of the swimming of microscopic organisms.
\newblock {\em Proc. Roy. Soc. A}, 209:447--461, 1951.

\bibitem{taylor52}
G.~I. Taylor.
\newblock The action of waving cylindrical tails in propelling microscopic
  organisms.
\newblock {\em Proc. Roy. Soc. A}, 211:225239, 1952.

\bibitem{gray55}
J.~Gray and G.~J. Hancock.
\newblock The propulsion of sea-urchin spermatozoa.
\newblock {\em J. Exp. Biol.}, 32:802--814, 1955.

\bibitem{lighthill76}
J.~Lighthill.
\newblock Flagellar hydrodynamics - {The John von Neumann} lecture, 1975.
\newblock {\em SIAM Rev.}, 18:161--230, 1976.

\bibitem{brennen77}
C.~Brennen and H.~Winet.
\newblock Fluid mechanics of propulsion by cilia and flagella.
\newblock {\em Ann. Rev. Fluid Mech.}, 9:339--398, 1977.

\bibitem{bergbook}
H.~C. Berg.
\newblock {\em {\it E. coli} in {M}otion}.
\newblock Springer-Verlag, NY, 2004.

\bibitem{braybook}
D.~Bray.
\newblock {\em Cell {M}ovements}.
\newblock Garland Publishing, New York, NY, 2000.

\bibitem{purcell77}
E.~M. Purcell.
\newblock Life at low {Reynolds} number.
\newblock {\em Am. J. Phys.}, 45:3--11, 1977.

\bibitem{becker03}
L.~E. Becker, S.~A. Koehler, and H.~A. Stone.
\newblock On self-propulsion of micro-machines at low {Reynolds} number:
  {Purcell's} three-link swimmer.
\newblock {\em J. Fluid Mech.}, 490:15 -- 35, 2003.

\bibitem{avron04:opt}
J.~E. Avron, O.~Gat, and O.~Kenneth.
\newblock Optimal swimming at low {Reynolds} numbers.
\newblock {\em Phys. Rev. Lett.}, 93:186001, 2004.

\bibitem{najafi04}
A.~Najafi and R.~Golestanian.
\newblock Simple swimmer at low {Reynolds} number: {Three} linked spheres.
\newblock {\em Phys. Rev. E}, 69:062901, 2004.

\bibitem{najafi05}
A.~Najafi and R.~Golestanian.
\newblock Propulsion at low {Reynolds} number.
\newblock {\em J. Phys. - Cond. Mat.}, 17:S1203--S1208, 2005.

\bibitem{avron05:pushme}
J.~E. Avron, O.~Kenneth, and D.~H. Oaknin.
\newblock Pushmepullyou: {A}n efficient micro-swimmer.
\newblock {\em New J. Phys.}, 7:234, 2005.

\bibitem{dreyfus05}
R.~Dreyfus, J.~Baudry, M.~L. Roper, M.~Fermigier, H.~A. Stone, and J.~Bibette.
\newblock Microscopic artificial swimmers.
\newblock {\em Nature}, 437:862--865, 2005.

\bibitem{WigginsGoldstein}
C.~H. Wiggins and R.~E. Goldstein.
\newblock Flexive and propulsive dynamics of elastica at low {Reynolds} number.
\newblock {\em Phys. Rev. Lett.}, 80:3879 -- 3882, 1998.

\bibitem{machin58}
K.~E. Machin.
\newblock Wave propagation along flagella.
\newblock {\em J. Exp. Biol}, 35:796--806, 1958.

\bibitem{machin63}
K.~E. Machin.
\newblock The control and synchronization of flagellar movement.
\newblock {\em Proc. Roy. Soc. B}, 158:88--104, 1963.

\bibitem{camalet99:prl}
S.~Camalet, F.~Julicher, and J.~Prost.
\newblock Self-organized beating and swimming of internally driven filaments.
\newblock {\em Phys. Rev. Lett.}, 82:1590--1593, 1999.

\bibitem{camalet00:njp}
S.~Camalet and F.~Julicher.
\newblock Generic aspects of axonemal beating.
\newblock {\em New J. Phys.}, 2:1--23, 2000.

\bibitem{gittes93}
F.~Gittes, B.~Mickey, J.~Nettleton, and J.~Howard.
\newblock Flexural rigidity of microtubules and actin-filaments measured from
  thermal fluctuations in shape.
\newblock {\em J. Cell Biol.}, 120:923--934, 1993.

\bibitem{Wiggins:Biophys}
C.~H. Wiggins, D.~Riveline, A.~Ott, and R.~E. Goldstein.
\newblock Trapping and wiggling: Elastohydrodynamics of driven microfilaments.
\newblock {\em Biophys. J.}, 74:1043--1060, 1998.

\bibitem{riveline97}
D.~Riveline, C.~H. Wiggins, R.~E. Goldstein, and A.~Ott.
\newblock Elastohydrodynamic study of actin filaments using fluorescence
  microscopy.
\newblock {\em Phys. Rev. E}, 56:R1330--R1333, 1997.

\bibitem{cebers05}
A.~Cebers.
\newblock Flexible magnetic filaments.
\newblock {\em Curr. Opin. Colloid Int. Sci.}, 10:167--175, 2005.

\bibitem{roper06}
M.~Roper, R.~Dreyfus, J.~Baudry, M.~Fermigier, J.~Bibette, and H.~A. Stone.
\newblock On the dynamics of magnetically driven elastic filaments.
\newblock {\em J. Fluid Mech.}, 554:167--190, 2006.

\bibitem{gauger06}
E.~Gauger and H.~Stark.
\newblock Numerical study of a microscopic artificial swimmer.
\newblock {\em Phys. Rev. E}, 74:021907, 2006.

\bibitem{manghi06}
M.~Manghi, X.~Schlagberger, and R.~R. Netz.
\newblock Propulsion with a rotating elastic nanorod.
\newblock {\em Phys. Rev. Lett.}, 96:068101, 2006.

\bibitem{powers02}
T.~R. Powers.
\newblock Role of body rotation in bacterial flagellar bundling.
\newblock {\em Phys. Rev. E}, 65:040903, 2002.

\bibitem{wolgemuth00}
C.~W. Wolgemuth, T.~R. Powers, and R.~E. Goldstein.
\newblock Twirling and whirling: {Viscous} dynamics of rotating elastic
  filaments.
\newblock {\em Phys. Rev. Lett.}, 84:1623--1626, 2000.

\bibitem{wada06}
H.~Wada and R.~R. Netz.
\newblock Non-equilibrium hydrodynamics of a rotating filament.
\newblock {\em Europhys. Lett.}, 75:645--651, 2006.

\bibitem{kim06:pumping}
Y.~W. Kim and R.~R. Netz.
\newblock Pumping fluids with periodically beating grafted elastic filaments.
\newblock {\em Phys. Rev. Lett.}, 96:158101, 2006.

\bibitem{lagomarsino03}
M.~C. Lagomarsino, F.~Capuani, and C.~P. Lowe.
\newblock A simulation study of the dynamics of a driven filament in an
  {Aristotelian} fluid.
\newblock {\em J. Theor. Biol.}, 224:215--224, 2003.

\bibitem{lowe03}
C.~P. Lowe.
\newblock Dynamics of filaments: {M}odelling the dynamics of driven
  microfilaments.
\newblock {\em Philos. Trans. R. Soc. London, Ser. B}, 358:1543--1550, 2003.

\bibitem{chwang71}
A.~T. Chwang and T.~Y. Wu.
\newblock Helical movement of micro-organisms.
\newblock {\em Proc. Roy. Soc. Lond. B}, 178:327--346, 1971.

\bibitem{keller76}
J.~B. Keller and S.~I. Rubinow.
\newblock Swimming of flagellated microorganisms.
\newblock {\em Biophys J.}, 16:151Ð170, 1976.

\bibitem{yu06}
T.~S. Yu, E.~Lauga, and A.~E. Hosoi.
\newblock Experimental investigations of elastic tail propulsion at low
  {Reynolds} number.
\newblock {\em Phys. Fluids}, 18:091701, 2006.

\bibitem{tillett70}
J.P.K. Tillett.
\newblock Axial and transverse {Stokes} flow past slender axisymmetric bodies.
\newblock {\em J. Fluid Mech.}, 44:401 -- 417, 1970.

\bibitem{batchelor70}
G.K. Batchelor.
\newblock Slender body theory for particles of arbitrary cross section in
  {Stokes} flow.
\newblock {\em J. Fluid Mech.}, 44:419 -- 440, 1970.

\bibitem{cox70}
R.~G. Cox.
\newblock The motion of long slender bodies in a viscous fluid. {P}art 1.
  {G}eneral theory.
\newblock {\em J. Fluid Mech.}, 44:791--810, 1970.

\bibitem{keller76-jfm}
J.~B. Keller and S.~I. Rubinow.
\newblock Slender body theory for slow viscous flow.
\newblock {\em J. Fluid Mech.}, 75:705--714, 1976.

\bibitem{geer76}
J.~Geer.
\newblock Stokes flow past a slender body of revolution.
\newblock {\em J. Fluid Mech.}, 78:577 -- 600, 1976.

\bibitem{johnson80}
R.~E. Johnson.
\newblock An improved slender body theory for {S}tokes flow.
\newblock {\em J. Fluid Mech.}, 99:411--431, 1980.

\bibitem{kimbook}
S.~Kim and J.~S. Karilla.
\newblock {\em Microhydrodynamics: {P}rinciples and {S}elected {A}pplications.}
\newblock Butterworth-Heinemann, Boston, MA, 1991.

\end{thebibliography}

\end{document}